\documentclass[prb,twocolumn,superscriptaddress, preprintnumbers,letterpaper,twoside]{revtex4}

\usepackage{times}
\usepackage{amsmath}
\usepackage{amsthm}
\usepackage{amssymb}
\usepackage{amsbsy}
\usepackage{amsfonts}
\usepackage{bm}
\usepackage{graphicx}
\usepackage{caption}
\usepackage{multirow}
\usepackage[all,pdftex,cmtip]{xy}
\usepackage{comment}
\usepackage[abs]{overpic}

\begin{document}


\title{Unexpected tunneling current from downstream neutral modes}

\author{Jennifer Cano}
\affiliation{Department of Physics, University of California, Santa Barbara,
California 93106, USA}
\author{Chetan Nayak}
\affiliation{Microsoft Research, Station Q, Elings Hall,
University of California, Santa Barbara, California 93106-6105, USA}
\affiliation{Department of Physics, University of California, Santa Barbara, California 93106, USA}

\begin{abstract}
We analyze transport through a quantum point contact in fractional quantum Hall states
with counter-propagating neutral edge modes.
We show that both the noise (as expected and previously calculated by other authors)
and (perhaps surprisingly) the average transmitted current are affected by downstream perturbations within the standard edge state model. We consider two different scenarios for downstream perturbations.
We argue that the change in transmitted current should be observable in experiments that have observed increased noise.
\end{abstract}

\maketitle

\section{Introduction}

The current traversing the edge of a quantum Hall device is elegantly described by chiral Luttinger liquid (CLL) theory.\cite{Wen92} For Laughlin states, the theory has only one edge mode, while for more complicated states, there might be multiple edge modes. Particle-hole conjugate states were originally predicted to have charged edge modes propagating in both directions\cite{Wen90,MacDonald90,MacDonald91}, but such counter-propagating charged modes were never detected.\cite{Ashoori92} This mystery was resolved when it was shown that in the presence of disorder and interactions, certain edges with counter-propagating charged modes could reconstruct into an edge with a single charged mode and counter-propagating neutral modes.\cite{Kane94a,Kane95} The question then remained, how can one detect the elusive neutral mode(s)? The question was answered by Bid, {\it et al.} \cite{Bid10}, who observed an increase in the noise across a quantum point contact (QPC) caused by a downstream perturbation, which they interpreted
as evidence for the existence of neutral excitations.

\begin{figure}[t]\centering
	\includegraphics[width=.48\textwidth]{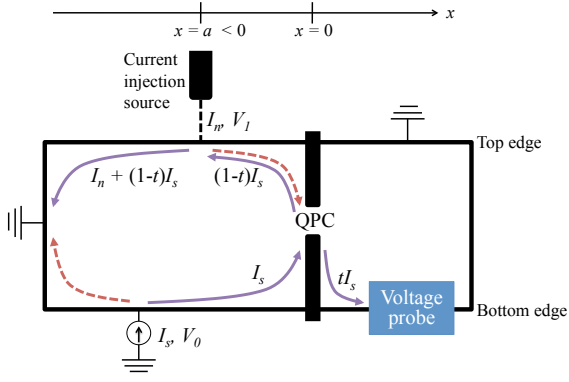}
	\caption{\textbf{Experimental set-up.} Current $I_n$ is injected into the top edge while current $I_s$ is sourced to the bottom edge. Solid purple lines show the direction of charged current, with magnitude indicated, while dotted red lines show the propagation of counterclockwise neutral modes.}
	\label{fig:ExptSetup}
\end{figure}

Measuring the shot noise across a QPC to confirm the $e/3$ charge of Laughlin's predicted quasiparticles was a breakthrough development in quantum Hall physics\cite{dePicciotto97,Saminadayar97,Kane94b,Fendley95}. Since then, significant effort has been devoted to using shot noise measurements to gain insight into more complicated edges.\cite{Reznikov99,Griffiths00,Comforti02,Dolev08,Bid09,Dolev11,Gross12,Inoue14a,Inoue14b,Rosenow02,Trauzettel04,Smits14,Feldman08} 
The experiment of Ref.~\onlinecite{Bid10} consists of a Hall bar with a QPC across which noise and current are measured, as shown in Fig~\ref{fig:ExptSetup}. Current is then injected into one edge, downstream of the QPC (here downstream always refers to the net direction of charged current), and the change in current and shot noise are measured. Intuitively, if the edge is chiral, then the current injection should not change the shot noise or current across the QPC; if the edge has a non-chiral charged mode, then the shot noise and current across the QPC should both change; if the edge has a non-chiral neutral mode, then shot noise across the QPC should increase but current should remain unchanged. Using this intuition, Ref~\onlinecite{Bid10} confirmed the existence of the counter-propagating neutral modes for $\nu = 2/3, 3/5$ and $5/2$, as well as confirmed the pure chirality of the edge at $\nu = 1/3, 2/5$ and $1$. This was a breakthrough experiment in understanding quantum Hall edge physics at particle-hole conjugate states.

At that time, a rigorous theoretical model of the experiment using CLL theory was absent. In trying to fill that void, we have found a surprising result that defies the intuitive prediction: a non-chiral neutral mode can change both the current and shot noise across the QPC. 

Our model assumes weak coupling between the quantum Hall edge and the external lead that injects current downstream of the QPC, which allows us to treat the effect of the current injection perturbatively. We first consider a toy model with fermionic edge modes and then move to a more general model with multiple Luttinger liquid edge modes that allows fractionalization and is expected to describe several Abelian particle-hole conjugate states. In both cases, we observe that injecting current downstream of the QPC changes the charged current across the QPC through the upstream propagation of neutral modes. The sign of the change depends on the scaling dimension of the tunneling quasiparticles.

We then consider the model proposed in Refs~\onlinecite{Takei11} and \onlinecite{Shtanko14}, which assumes that the effect of injecting current into an edge is to increase the temperature of that edge. We show that the increased temperature also changes the tunneling current across the QPC.

Finally, we compare the theoretical models to experimental results at $\nu = 2/3$. Both models predict a decrease in the magnitude of the tunneling current, which could reach tenths of nanoamps over the parameter regime of the experiment. Given the precision of the experiment, we believe this to be an observable effect. We then discuss directions for future work.

\section{Weak Downstream Perturbations}
\label{sec:weak-tunneling}

The experiment has two main features that need to be included in our model: a QPC at $x=0$, across which current and voltage are measured, and a current source at $x=a$ that injects current into the quantum Hall edge, downstream of the QPC; a schematic of the experimental set-up is shown in Fig~\ref{fig:ExptSetup}. The QPC is modeled in the usual way, by including terms which tunnel quasiparticles between the edges of the Hall bar.\cite{Kane92a,Kane92b} We model the downstream current source similarly: we assume the current source is a metallic lead which is weakly coupled to the quantum Hall edge and tunnels electrons between the two. We first compute the tunneling current and shot noise across the QPC in the absence of the injected current. We then turn on the current injection and compute the change in current and shot noise. If the edge is completely chiral, then the injected current, which enters downstream of the QPC, has no effect on the shot noise and current across the QPC. Here, we are interested in the more interesting case in which the edge consists of a chiral charged mode and an anti-chiral neutral mode; this can result from the equilibration of two counter-propagating charged modes, as in the particle-hole conjugates of the Laughlin states. We warm up by considering a toy model of free fermion edge modes and then generalize to an arbitrary edge, which permits fractionally charged excitations.

\subsection{Free fermion edge}
\label{sec:weak-tunn-fermion}

The edge of our free fermion model consists of a counterclockwise-propagating charged mode and a clockwise-propagating neutral mode. We denote the fermion annihilation operators for these modes $\psi_c$ and $\psi_n$, respectively. Since the `top' (T) and `bottom' (B) edges, as indicated in Fig~\ref{fig:ExptSetup}, are separated by grounded contacts, each annihilation operator has a subscript $T/B$ and the Lagrangian is a sum of separate Lagrangians on each edge $\mathcal{L} = \mathcal{L}_T + \mathcal{L}_B$, where
\begin{align}\mathcal{L}_{T/B} = \frac{1}{2\pi}\int dx & \left( \psi_{c,T/B}^\dagger(\pm \partial_t + v_{c,T/B}\partial_x)\psi_{c,T/B} \right. \nonumber\\
&\left. + \psi_{n,T/B}^\dagger(\mp \partial_t + v_{n,T/B} \partial_x ) \psi_{n,T/B}\right) \label{eq:free-fermion}\end{align}
The signs $\pm,\mp$ correspond to the $T/B$ edge and $v_{c/n,T/B}$ is the velocity of the indicated mode. The QPC at $x=0$ is incorporated in the Lagrangian through the tunneling term
\begin{equation} \mathcal{L}_{tun}^{1} = -\lambda_{1} \psi^\dagger_{c,T}\psi_{c,B}\delta(x) + h.c. \label{eq:L1-tun}\end{equation}
There is also a pair-tunneling term that mixes the charged and neutral modes:
\begin{align} \mathcal{L}_{tun}^{2} &= -\lambda_{2}\psi^\dagger_{c,T}\psi_{n,T}^\dagger\psi_{c,B}\psi_{n,B}\delta(x) + h.c. \label{eq:L2-tun}
 \end{align}
Similar single- and pair-tunneling terms involving only the neutral modes may also be present, but they do not contribute to the shot noise or current. Terms involving more fields or derivatives might also be present, but we do not need to consider them here.
We will usually consider an applied voltage, $V_0$, across the QPC, which is incorporated through the phase of the tunneling coefficients: $\lambda_{1,2} \rightarrow \lambda_{1,2}e^{-i \omega_0t}$, where $\omega_0 = eV_0$. 

We model the current injection by a second QPC at $x=a$ that allows tunneling from an external lead, whose fermion annihilation operator we denote by $\Psi$. We will always take $a<0$, as shown in Fig~\ref{fig:ExptSetup}. Single electron tunneling across the QPC is included by the term:
\begin{equation} \mathcal{L}_{inj}^1 = -\Lambda_1 \Psi^\dagger \psi_{c,T}\delta(x-a) + h.c.  \end{equation}
Now suppose that the lead also has a neutral fermion mode, whose annihilation operator is $\Psi_n$. Then there is also a pair tunneling term that mixes the charged and neutral fermion modes:
\begin{equation} \mathcal{L}_{inj}^{2} = -\Lambda_2 \Psi^\dagger\Psi_n^\dagger \psi_{c,T} \psi_{n,T}\delta(x-a)  + h.c.\label{eq:L2-inj} \end{equation}
The voltage difference $V_1$ between the external lead and the top edge of the Hall bar is incorporated through the tunneling coefficients by $\Lambda_{1,2} \rightarrow \Lambda_{1,2}e^{-i\omega_1 t}$, where $\omega_1 = eV_1$.

The charged current operator across the QPC at $x=0$ is given by $I_{tun} = e\frac{d}{dt}\langle N \rangle = -ie\langle \left[ N, H \right] \rangle = I_1 + I_2$, where $N = \psi_{c,T}^\dagger\psi_{c,T}$ is the electron number of the charged field and
\begin{align} I_{1} &=  -ie\lambda_{1}e^{-i\omega_0 t}\psi_{c,T}^\dagger\psi_{c,B}\delta(x)  + h.c. \\
I_{2} &=  -ie\lambda_{2}e^{-i\omega_0t}\psi^\dagger_{c,T}\psi^\dagger_{n,T}\psi_{c,B}\psi_{n,B}\delta(x) +h.c. 
\end{align}
Similarly, the operator that measures current across the QPC at $x=a$ is given by $I_{inj} = I_{\Psi 1} + I_{\Psi 2}$, where
\begin{align}I_{\Psi 1} &= -ie\Lambda_1e^{-i\omega_1 t}\Psi^\dagger\psi_{c,T}\delta(x-a)+h.c. \\
I_{\Psi 2} &= -ie\Lambda_2 e^{-i\omega_1 t}\Psi^\dagger\Psi_n^\dagger \psi_{c,T}\psi_{n,T}\delta(x-a) + h.c. 
\end{align}
It is straightforward to compute $\langle I_{tun} \rangle_0$ and $\langle I_{inj} \rangle_0$, where the subscript 0 indicates the lowest order in perturbation theory:
\begin{align}
\langle I_{tun}\rangle_0 &= 2\pi e \omega_0|\lambda_{1}|^2 + \frac{\pi e}{3}\omega_0^3 |\lambda_{2}|^2\label{eq:I-tun-0} \\
\langle I_{inj}\rangle_0 &= 2\pi e \omega_1 |\Lambda_1|^2 +  \frac{\pi e}{3}\omega_1^3|\Lambda_2|^2 \label{eq:I-inj-0}
\end{align}
We have set $v_{i,T/B}=1$ by absorbing them in the tunneling coefficients.
We now show that to the next order in perturbation theory, $\langle I_{tun} \rangle$ depends on the injected current at $x=a$, \textit{even though only neutral excitations move the towards the QPC at $x=0$}. Specifically, we define
\begin{equation} \Delta I_{tun} \equiv  \langle I_{tun} \rangle - \langle I_{tun} \rangle |_{\Lambda_{i}=0} \label{eq:delta-Itunn-def}\end{equation}
and show that $\Delta I_{tun} \neq 0$. It is not hard to show that $I_1$ is independent of the current injected downstream (and hence independent of $\Lambda_{1,2}$) because it depends only on the charged edge mode, which moves from $x=a$ to the left ground without passing the QPC at $x=0$. Hence,
\begin{align} &\Delta I_{tun} = \langle I_{2} \rangle - \langle I_{2} \rangle |_{\Lambda_{i} = 0}\nonumber\\ 
&= \langle I_{2} \left( i\int dt_1 \mathcal{L}_{tun}^{2} \right)\left( i\int dt_2 \mathcal{L}_{inj}^{2} \right) \left(  i\int dt_3\mathcal{L}_{inj}^{2}\right) \rangle_0 \nonumber\\
&\quad\quad+ \mathcal{O}(|\lambda_{2}|^2 |\Lambda_2|^4,|\lambda_{2}|^4 |\Lambda_2|^2) \label{eq:define-corr-function}
\end{align} 
Because the system is not in equilibrium at non-zero voltage, the correlation function in Eq~(\ref{eq:define-corr-function}) requires careful treatment, which is explained in Appendix~\ref{sec:Keldysh}. The full calculation is shown in Appendix~\ref{sec:fermion-calc}. Here we state the result:
\begin{equation}
\Delta I_{tun} = \frac{4\pi^3}{3}e|\lambda_{2}|^2|\Lambda_2|^2\omega_0 \omega_1^4 \label{eq:tunn-current-fermions} 
\end{equation}
We have assumed that the neutral fermions are Majorana fermions, which yields the physically reasonable result that $\Delta I_{tun} = 0$ when $\omega_0=0$. If this is not the case, then $\Delta I_{tun}$ will have additional terms which are odd in $\omega_1$. These terms are computed in Appendix~\ref{sec:fermion-calc}.

Eq~(\ref{eq:tunn-current-fermions}) shows the main point of this work and gives an experimental prediction for this fictitious edge: the charged current measured across the QPC at $x=0$ will change when current is injected at $x=a$, even though it is only carried to $x=0$ by the neutral mode. Physically, we understand this as the presence of extra neutral fermions enhancing the probability of a pair-tunneling event. Since tunneling events in one direction are favored to begin with, due to the voltage bias, the probability for these events is more enhanced, leading to increased current. That this is realized in such a simple model hints that it is a general result, which applies to any edge with oppositely propagating neutral and charged modes and tunneling operators that mix the two. 

\subsection{Luttinger liquid edge}
\label{sec:weak-tunneling-bosons}

To generalize the results of the previous section to edges with fractional excitations, we describe the edge by a Luttinger liquid with counter-propagating charged and neutral modes, denoted by the bosonic fields $\phi_c$ and $\phi_n$. We assume the Lagrangian is diagonal in these modes after scattering and interactions have been included. In the set-up shown in Fig~\ref{fig:ExptSetup}, the Lagrangian is a sum of Lagrangians on the top and bottom edges, $\mathcal{L} = \mathcal{L}_T + \mathcal{L}_B$, where
\begin{align} \mathcal{L}_{T/B} = &\frac{1}{4\pi}\left[ \int dx \, g_c\left( \pm  \partial_t + v_{c,T/B} \partial_x\right)\phi_{c,T/B}\partial_x\phi_{c,T/B}\right. \nonumber\\
& + g_n \left( \mp \partial_t + v_{n,T/B} \partial_x\right)\phi_{n,T/B}\partial_x\phi_{n,T/B} \Bigg]
\label{eq:bosonic-lagrangian} \end{align}
where $T/B$ denotes the top or bottom edge, the $v$'s denotes the velocities, and $g_c$ and $g_n$ are integers that determine the scaling dimensions of operators in the theory. 
A quasiparticle is labelled by an integer pair $q=(q_n,q_c)$, which determines its annihilation operator, $\Phi_q = e^{iq_n\phi_n + iq_c\phi_c}$ and its charge, $q_c e^*$, where $e^*$ is the minimum quasiparticle charge. By convention, we take $q_c>0$; the hermitian conjugate terms correspond to creation operators. The scaling dimension of $\Phi_q$ is given by $q^2/2$, where we have defined the inner product $q^2 = q\cdot q \equiv q_c^2/g_c + q_n^2/g_n$. For a particular edge theory, not all pairs are allowed excitations; for example, for the $\nu = 2/3$ edge that we will discuss in more detail in Sec~\ref{sec:comparison},  $g_c = 6$, $g_n = 2$, $e^* = 1/3$ and allowed excitations have $q_c = q_n \mod 2$.\cite{Kane94a}

The QPC at $x=0$ is included in the Lagrangian through the tunneling term,
\begin{equation} \mathcal{L}_{tun} = -\sum_{q,q'} \lambda_{qq'} \Phi_{q,T}^\dagger \Phi_{q',B} \delta(x) + h.c. \label{eq:L-tun-boson} \end{equation}
For $\lambda_{qq'}$ to be nonzero, the quasiparticles $q$ and $q'$ must have the same charge and statistics, but not necessarily the same neutral component. 
If there is a voltage $V_0$ across the QPC then $\lambda_q \rightarrow \lambda_q e^{-iq_c \omega_0 t}$, where $\omega_0 = e^* V_0$. The current injection is described by a QPC at $x=a$ that tunnels electrons between an external lead and the quantum Hall edge. Let $\Psi$ denote the electron annihilation operator of the lead. Then the current injection is described by the Lagrangian,
\begin{equation} \mathcal{L}_{inj} = -  \sum_{ r } \Lambda_r  \Psi^\dagger \Phi_{r,T} \delta(x-a) + h.c. \label{eq:L-inj-boson}\end{equation}
where $\Lambda_r$ is only nonzero if $r_ce^* = e$, so that the sum is over all electron operators in the theory.
If there is a voltage $V_1$ across the QPC at $x=a$ then $\Lambda_r \rightarrow \Lambda_r e^{-i\omega_1 t}$, where $\omega_1 = eV_1$. Less relevant terms that tunnel pairs of electrons across the QPC might also be present, but we do not consider them here.

The tunneling current operators are given by $I_{tun} = \sum_{q,q'} I_{qq'}$ and $I_{inj} = \sum_{r} I_{r\Psi}$, where
\begin{align}
I_{qq'} & = -iq_ce^*\lambda_{qq'} \Phi_{q,T}^\dagger \Phi_{q',B} \delta(x) + h.c. \\
I_{r\Psi} &= -ie\Lambda_r \Psi^\dagger \Phi_{r,T} \delta(x-a) + h.c. \end{align}
It is straightforward to compute to lowest order in perturbation theory, setting $v_{i,T/B}=1$ by absorbing the velocities into the tunneling coefficients,
\begin{align}
\langle I_{tun} \rangle_0 &= \sum_{q,q' } \frac{2\pi e^*q_c}{\Gamma(q^2+ q'^2)}|\lambda_{qq'}|^2 \text{sgn}(\omega_0) |q_c \omega_0|^{q^2+ q'^2-1}\label{eq:I-tun-boson-leading}\\
\langle I_{inj} \rangle_0 &=  \sum_{r } \frac{2\pi e }{\Gamma(1+r^2)}|\Lambda_r|^2\text{sgn}(\omega_1)|\omega_1|^{r^2} \label{eq:I-inj-boson-leading}\end{align}
These results are identical to Eqs~(\ref{eq:I-tun-0}) and (\ref{eq:I-inj-0}) when the scaling dimensions of the tunneling terms are matched.   The zeroth order, zero-frequency shot noise is given by the sum 
\begin{equation} S(\omega=0) = \sum_{ q,q'  } q_c e^* \left| \langle I_{qq'} \rangle_0 \right| \end{equation} 
As in the previous section, we want to find the change in tunneling current at $x=0$ in the presence of the injection at $x=a$, when $a<0$, so the current moving from the injection to the QPC at $x=0$ is carried only by the neutral mode. We define the change in current by $\Delta I_{tun}$ in Eq~(\ref{eq:delta-Itunn-def}). To leading order, 
\begin{align} \Delta I_{tun} &=\langle I_{tun} \left( i\int dt_1 \mathcal{L}_{tun} \right)\left( i\int dt_2 \mathcal{L}_{inj} \right) \left(  i\int dt_3\mathcal{L}_{inj}\right) \rangle_0 
\label{eq:define-corr-function-boson}
\end{align} 
To ensure that $\Delta I_{tun} = 0$ when $\omega_0=0$, we assume that the tunneling coefficient $\lambda_{qq'}$ for tunneling a quasiparticle with $q=(q_n,q_c)$ from the top edge is the same as that for a quasiparticle with opposite neutral charge, $q=(-q_n,q_c)$, and similarly for tunneling $q'$ from the bottom edge and $r$ to the external lead $\Lambda_r$.\footnote{This assumption is consistent with the experimental observation\cite{Bid10} that excess shot noise is symmetric under reversing the sign of the source current $I_s$. An observation of excess shot noise not symmetric under $I_s \rightarrow -I_s$ would demonstrate that quasiparticles with opposite neutral charge $q_n$ do not have equal tunneling amplitudes.}
Here we state the result in two limiting cases; the full expression is given in Appendix~\ref{sec:boson-calc}.
In the limit $|\omega_1|\ll |\omega_0|$,
\begin{equation} \frac{\Delta I_{tun}}{e^*\text{sgn}(\omega_0)} = \sum_{\substack{q,q',r \\ r_n,q_n\neq 0}}  b_{qq',r} \frac{|\omega_1|^{r^2+1}|\omega_0q_c|^{q^2 + q'^2-3}}{\Gamma(r^2+2)\Gamma(q^2 + q'^2-2)}\label{eq:tunn-current-bosons-2}\end{equation}
while in the limit $|\omega_1|\gg |\omega_0|$,
\begin{align} \frac{\Delta I_{tun}}{e^*\text{sgn}(\omega_0)} &= \sum_{\substack{q,q', r \\ r_n,q_n\neq 0\\ 0<q^2+q'^2<2}}  -b_{qq',r}  \frac{|\omega_1|^{r^2-1}|\omega_0q_c|^{q^2 + q'^2-1}}{\Gamma(r^2)\Gamma(q^2 + q'^2)}  \nonumber\\
&+  \sum_{\substack{q,q', r \\ r_n,q_n\neq 0\\q^2+q'^2>2}}  b_{qq',r}  \frac{|\omega_1|^{r^2+q^2 + q'^2-3}|\omega_0q_c|}{\Gamma(r^2+q^2 + q'^2-2)} 
\label{eq:tunn-current-bosons}\end{align}
where $b_{qq',r}\propto q_c |\lambda_{qq'}|^2|\Lambda_r|^2$ is a positive constant.
Eqs~(\ref{eq:tunn-current-bosons-2}) and (\ref{eq:tunn-current-bosons}) agree with Eq~(\ref{eq:tunn-current-fermions}) after identifying $r^2=3, q'^2+q^2=4$ and show our main result in general form: \textit{electrons injected into the edge at $x=a$ will cause a change in the charged tunneling current at $x=0$, even though only the neutral part of the injected electrons move from $x=a$ to $x=0$.} This general formulation is applicable to any bosonized Abelian quantum Hall edge with counter-propagating modes. We expect it could be extended to a non-Abelian edge by matching the scaling dimensions of the tunneling operators. Interestingly, though, the sign of $\Delta I_{tun}$ depends on the magnitude of the scaling dimensions $q$ and $q'$ and the simple picture of enhanced tunneling in the fermionic model is generalized to enhanced or diminished tunneling depending on the scaling dimensions of the tunneling quasiparticles.

When we compare to experimental data we will want the excess zero-frequency shot noise, which is also computed in Appendix~\ref{sec:boson-calc}. We find that in the limit $|\omega_1|\ll |\omega_0|$,
\begin{equation}\Delta S_{tun}=  (e^*)^2 \sum_{\substack{q,q', r \\ r_n,q_n\neq 0 }} q_c b_{qq',r} \frac{|\omega_1|^{r^2+1}|\omega_0q_c|^{q^2 + q'^2-3}}{\Gamma(r^2+2)\Gamma(q^2 + q'^2-2)}\label{eq:excess-noise-bosons-2}\end{equation}
while in the limit $|\omega_1|\gg |\omega_0|$,
\begin{align} \Delta S_{tun} &= (e^*)^2  \sum_{\substack{q,q', r \\ r_n,q_n\neq 0}}  \frac{q_c b_{qq',r} |\omega_1|^{r^2+q^2 + q'^2-2}}{\Gamma(r^2+q^2 + q'^2-1)}
\label{eq:excess-noise-bosons}\end{align}

\section{Temperature difference model}
\label{sec:strong-tunneling}

In the previous section, we modeled the current injection by an external lead weakly coupled to the edge of the Hall bar by a QPC. An alternate approach is used in Refs~\onlinecite{Takei11} and \onlinecite{Shtanko14}. There, they assume that the sole effect of the current injection is to increase the temperature of the top edge of the Hall bar, while the temperature of the bottom edge remains constant. The increase in temperature is responsible for an increase in shot noise across the QPC at $x=0$, which is computed in Refs~\onlinecite{Takei11} and \onlinecite{Shtanko14}. However, it does not appear to have been noted that the increase in temperature also changes the magnitude of the tunneling current. Here, we write an expression for $\Delta I_{tun}$ when there is a temperature difference between the two edges of the Hall bar and describe how it differs qualitatively from our prediction for $\Delta I_{tun}$ in the previous section.

We again describe the edge of the quantum Hall bar using Luttinger liquid theory. We consider an edge with counter-propagating charged and neutral modes, described by the Lagrangian $\mathcal{L} = \mathcal{L}_T + \mathcal{L}_B$, where $\mathcal{L}_{T/B}$ are given by Eq~(\ref{eq:bosonic-lagrangian}). 
Quasiparticles are labelled by integer pairs $q=(q_n,q_c)$, with their charge given by $q_c e^*$ and annihilation operator $\Phi _m$. 
The QPC at $x=0$ is described by $\mathcal{L}_{tun}$ in Eq~(\ref{eq:L-tun-boson}). When there is a voltage $V_0$ applied across the QPC, $\lambda_q \rightarrow \lambda_q e^{-iq_c\omega_0t}$, where $\omega_0 = e^* V_0$. Following Ref~\onlinecite{Shtanko14}, the tunneling current from a particular species of quasiparticle $\Phi_q$ from the top edge of the Hall bar to a species $\Phi_{q'}$ on the bottom edge can be computed when the top edge is at temperature $T_T$ and the bottom edge at $T_B$ using the finite temperature prescription for correlation functions described in Sec~\ref{sec:finite-temp}, yielding 
\begin{align} \langle I_{tun} \rangle_0 \! &= \! \text{sgn}(\omega_0) 4 \sum_{ q,q'  }  q_ce^*|\lambda_{qq'} |^2 (\pi T_B)^{q^2 + q'^2-1} \nonumber\\
&\times  \sin(\frac{\pi}{2} (q^2+q'^2)) F\left(\frac{q_c |\omega_0|}{\pi T_B}, \frac{T_T}{T_B}\right) \label{eq:tunn-current-strong}\end{align}
where $F$ is the integral
\begin{equation} F(\alpha,\beta) \!=\! \int_0^\infty dx   \frac{\beta^{q^2}\sin(\alpha x)}{(\sinh(\beta x))^{q^2}(\sinh(x))^{q'^2}} \label{eq:strong-integral}\end{equation}
We have absorbed the edge mode velocities into the tunneling coefficients. When there are multiple species of quasiparticles, their contributions to the tunneling current add.

Define the excess current
$\Delta I_{tun} = \langle I_{tun} \rangle - \langle I_{tun}\rangle  _{T_T=T_B}$. The contribution to $\Delta I_{tun}$ from tunneling from $\Phi_q$ to $\Phi_{q'}$, transferring charge $q_c e^*$, is proportional to 
\begin{equation} F\left(\frac{q_c |\omega_0|}{\pi T_B}, \frac{T_T}{T_B}\right) - F\left(\frac{q_c |\omega_0|}{\pi T_B}, 1 \right) < 0\end{equation}
where the inequality results from imposing the physical constraint $T_T/T_B>1$. Consequently, when $\sin(\frac{\pi}{2}( q^2+q'^2))>0$, increasing the injected current decreases the magnitude of the tunneling current, while in the opposite case, the magnitude of the tunneling current decreases. As in the previous section, the sign of the change in current depends on the scaling dimensions of the tunneling quasiparticles.



To fit the temperature difference theory to the experimental data in the next section, we will need the expression for zero-frequency noise measured at the voltage probe when $I_n = 0$, which is computed in Ref~\onlinecite{Shtanko14}. The contribution to the noise from the process of tunneling $\Phi_q$ on the top edge to $\Phi_{q'}$ on the bottom edge is
\begin{align} &S_{tun}|_{T_T = T_B} = -\frac{2}{\pi}(2\pi T_B)^{2q^2+2q'^2-1}(q_c e^*)^2 |\lambda_{qq'}|^2\nonumber\\
&\times \sinh\left( \frac{1}{2}\pi \alpha_q \right)B(q^2 + q'^2 +\frac{i\alpha_q}{2} ,q^2 + q'^2-\frac{i\alpha_q}{2} ) \nonumber\\
&\times  i\left( \psi(q^2 + q'^2+\frac{i\alpha_q}{2} ) -  \psi(q^2 + q'^2-\frac{i\alpha_q}{2} )\right)
\label{eq:excess-noise-strong}\end{align}
where $\alpha_q = q_c\omega_0 /(\pi T_B)$, $B$ is the beta function and $\psi$ is the digamma function.
The contributions from multiple species of quasiparticles add.
Note that there is an important diference between this model and the model of a weak downstream perturbation in Section \ref{sec:weak-tunneling}. If the two edges are at different temperatures, then all tunneling processes are affected by the temperature difference. However, in the case of a weak downstream perturbation, only tunneling processes involving counter-propagating neutral modes will be affected. In the case of the $\nu=2/3$ state, these would be charge-$e/3$ tunneling processes; charge-$2e/3$ tunneling processes, which do not involve the neutral modes, would be independent of the downstream perturbation in the limit of vanishing interaction between charged and neutral modes. 

\section{Comparison to experiment at $\nu = 2/3$}\label{sec:comparison}

We have shown that injecting current downstream from the QPC should produce a change in the tunneling current across the QPC. In this section, we try to estimate the magnitude of this change when the system is at filling fraction $\nu = 2/3$ to determine whether it could be observed in experiment. When we model the current injection by weakly coupling a lead to the edge of the Hall bar, as described in Sec~\ref{sec:weak-tunneling}, we do this by fitting our theoretical expression for $\Delta S_{tun}$ to the measured excess shot noise in Ref~\onlinecite{Bid10}. The best-fit values of the tunneling amplitudes allow us to estimate $\Delta I_{tun}$. In the temperature difference model of Sec~\ref{sec:strong-tunneling} we fit the measured shot noise at $I_n=0$ from Ref~\onlinecite{Bid10} and use best-fit parameters from Ref~\onlinecite{Shtanko14} to estimate $\Delta I_{tun}$.
Coincidentally, in both models we find that there should be tunneling current on the order of .1 nA, which should be observable in experiment.

\subsection{Theoretical description of the $\nu = 2/3$ edge}

The $\nu = 2/3$ edge is expected to be described by the Lagrangian~(\ref{eq:bosonic-lagrangian}), with $g_c = 6, g_n = 2$ and $e^* = e/3$.\cite{Kane94a} Tunneling across the QPC at $x=0$, as described by Eq (\ref{eq:L-tun-boson}), is dominated by three equally most-relevant terms. Two of these terms tunnel charge $e/3$ quasiparticles, described by $q_n=\pm 1, q_c = 1$ and the third tunnels a charge $2e/3$ quasiparticle with $q_n = 0, q_c = 2$. Other quasiparticles are less relevant and we will not consider them here. Either species of charge-$e/3$ quasiparticle can tunnel from the top edge of the Hall bar to either species at the bottom edge; let $|\lambda_1|^2$ denote the sum of the squares of the amplitudes corresponding to charge $e/3$ tunneling across the Hall bar and let $|\lambda_2|^2$ denote the square of the amplitude corresponding to tunneling charge $2e/3$ across the Hall bar. 
Tunneling from the external electron lead, described in Eq~(\ref{eq:L-inj-boson}), is dominated by two most-relevant terms, which have $r_n = \pm 1, r_c = 3$. We denote their respective couplings $\Lambda_1,\Lambda_2$. Hence, all of these most-relevant tunneling terms have $q^2 = 2/3$ and $r^2=2$.

 The experimental data is in terms of the source current, $I_s$, and the injected current, $I_n$, which we need to express in terms of our theoretical parameters $V_0$ and $V_1$ (weakly coupled lead) or $V_0, T_T$ and $T_B$ (temperature difference model). The source current is related to the voltage $V_0$ applied across the Hall bar by the Hall conductance, $I_s = \frac{2}{3}\frac{e^2}{h}V_0$. In the weak coupling case, Eq~(\ref{eq:I-inj-boson-leading}) yields $I_n \equiv \langle I_{inj}\rangle_0 \propto \text{sgn}(\omega_1)\omega_1^2$, where $\omega_1 = eV_1$. In the temperature difference model, we use the fit from Ref~\onlinecite{Shtanko14}, which expresses the temperatures in terms of the injected current by $T_T/T_B - 1 \propto |I_n|^p $, where $p$ is determined from the fit.

\subsection{Theoretical prediction of excess current for the $\nu = 2/3$ edge weakly coupled to the current injection}

We fit our theoretical expression for $\Delta S_{tun}$ to data in Ref~\onlinecite{Bid10} and use the fit to predict $\Delta I_{tun}$. The experimental data includes a measurement of excess shot noise as a function of $I_n$ when $I_s = 0$ and as a function of $I_s$ for several values of $I_n$. 

\begin{figure}[h]\centering
	\includegraphics[width=.48\textwidth]{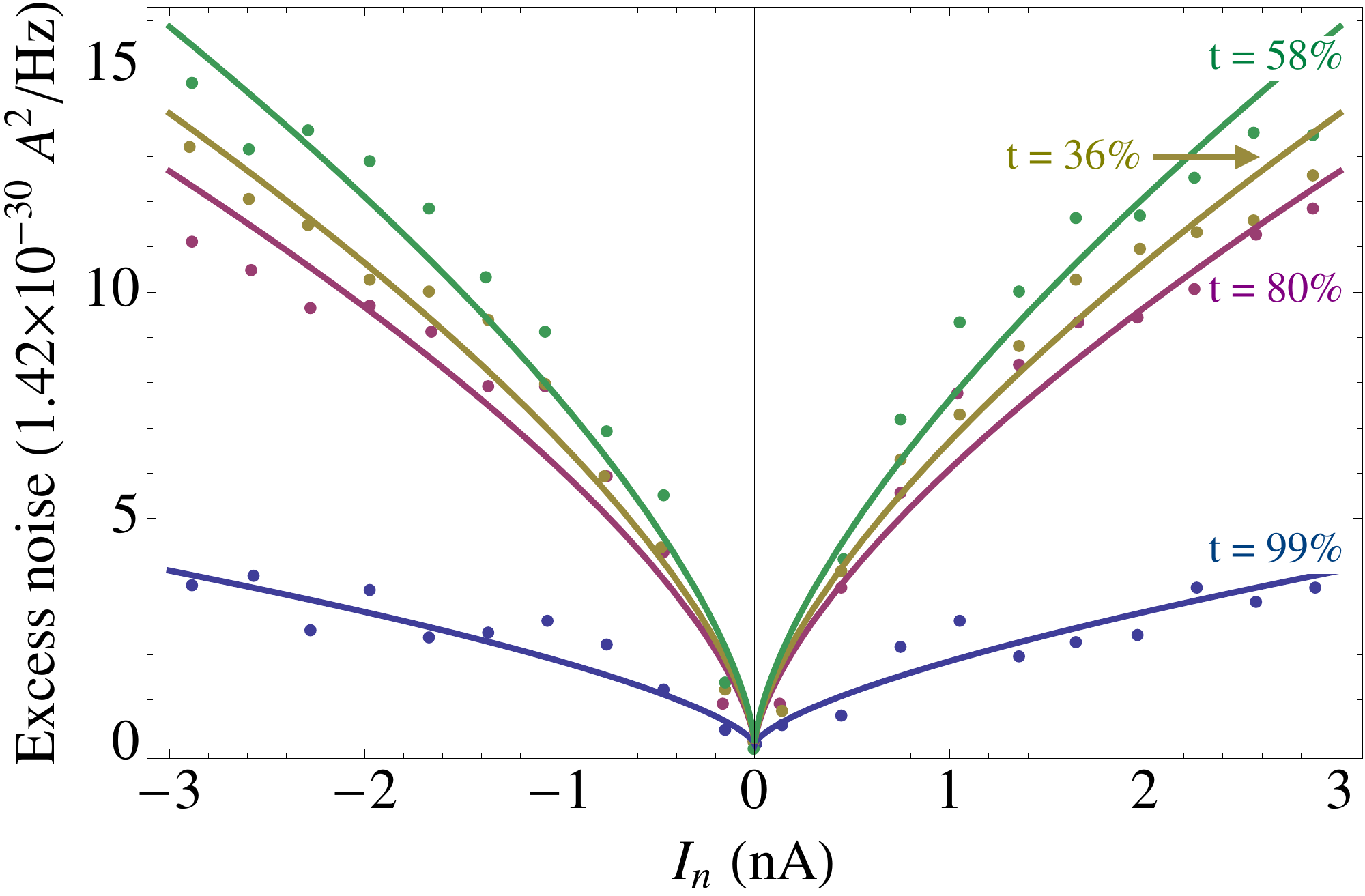}
	\caption{\textbf{Excess shot noise as a function of $I_n$ at $\nu = 2/3$}. Dots show experimental data at several transmission probabilities $t$ and lines indicate the theoretical prediction $\Delta S_{tun}\propto |I_n|^{2/3}$.}
	\label{fig:BidDataFig2}
\end{figure}

The excess shot noise as a function of $I_n$ when $I_s = \omega_0=0$ is shown in Fig~\ref{fig:BidDataFig2}, overlaid with the experimental data for several transmission probabilities $t$ from Fig 2 in Ref~\onlinecite{Bid10}. Using Eq~(\ref{eq:excess-noise-bosons}), our theory predicts the scaling $\Delta S_{tun} \propto |\omega_1|^{4/3} \propto |I_n|^{2/3}$, which is plotted with only an overall scaling factor for each $t$. We have taken $T=0$ for simplicity. The theoretical model fits the experimental data well at all transmission probabilities. It is especially good at $t=99\%$, where perturbation theory is most applicable.

\begin{figure*}
        \centering
        \includegraphics[width=8.4cm]{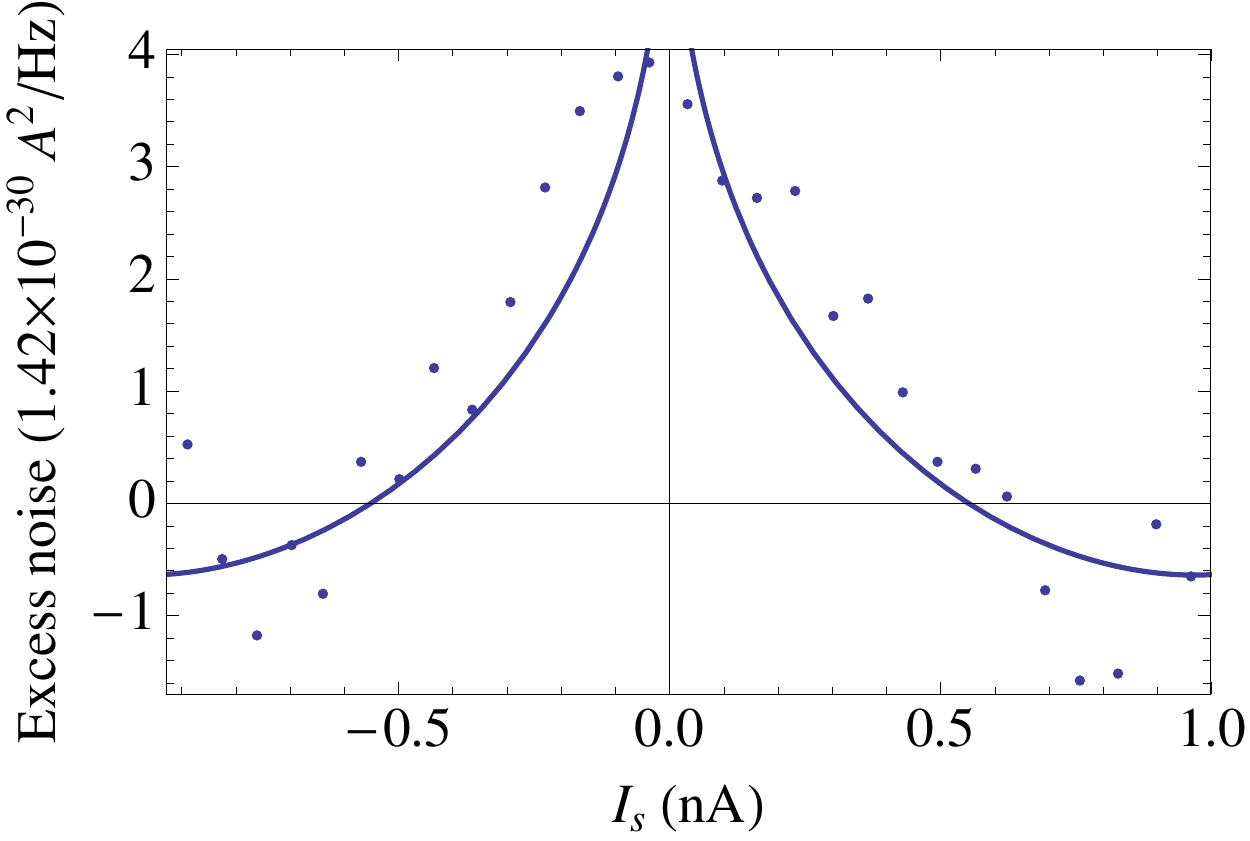}  \includegraphics[width=8.4cm]{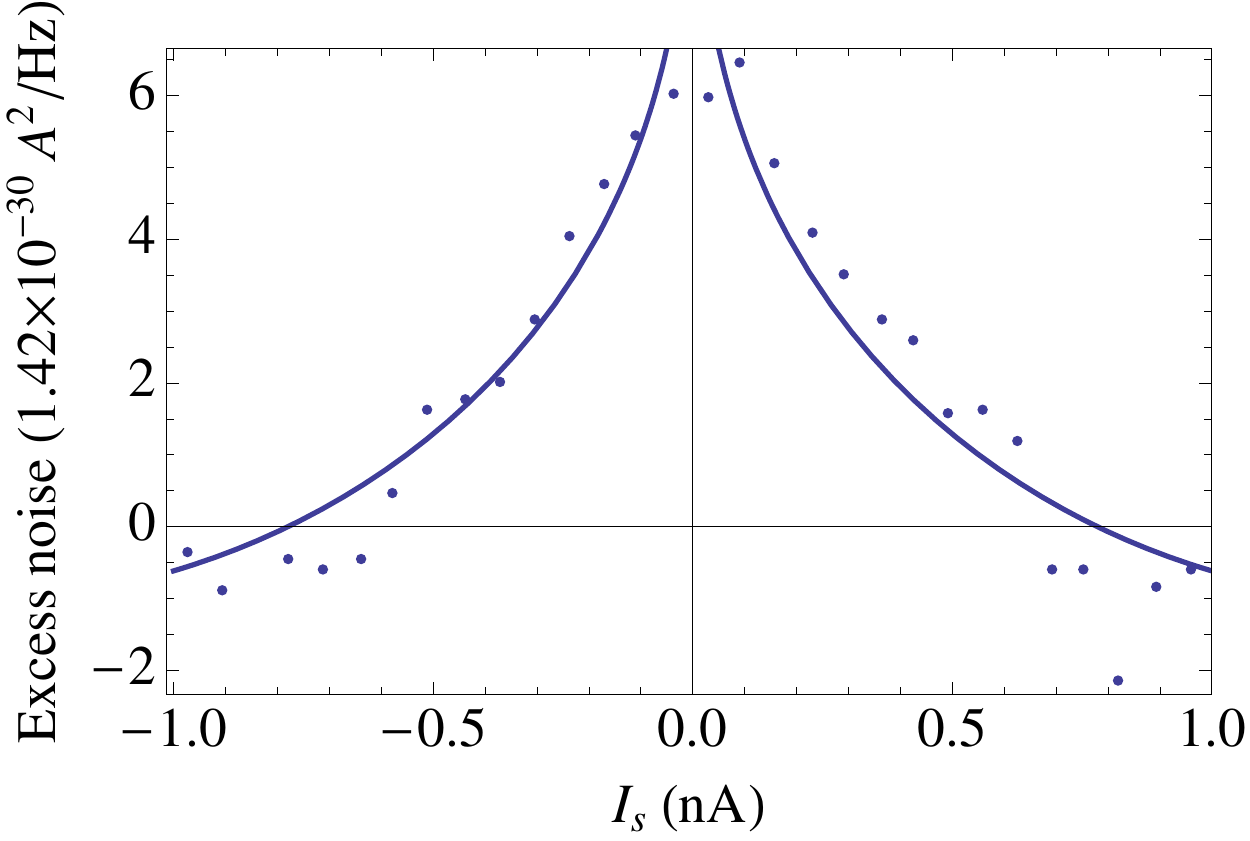}\\
        \includegraphics[width=8.4cm]{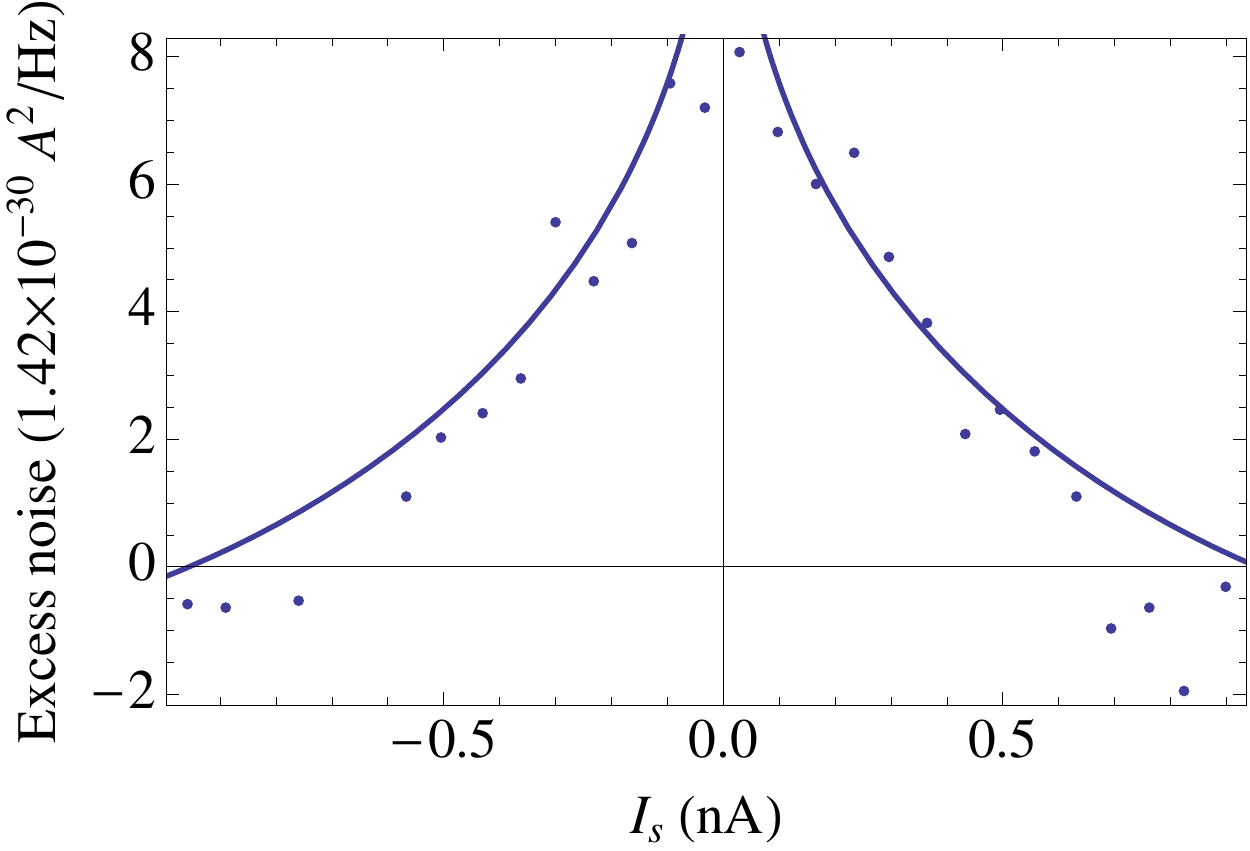}  \includegraphics[width=8.4cm]{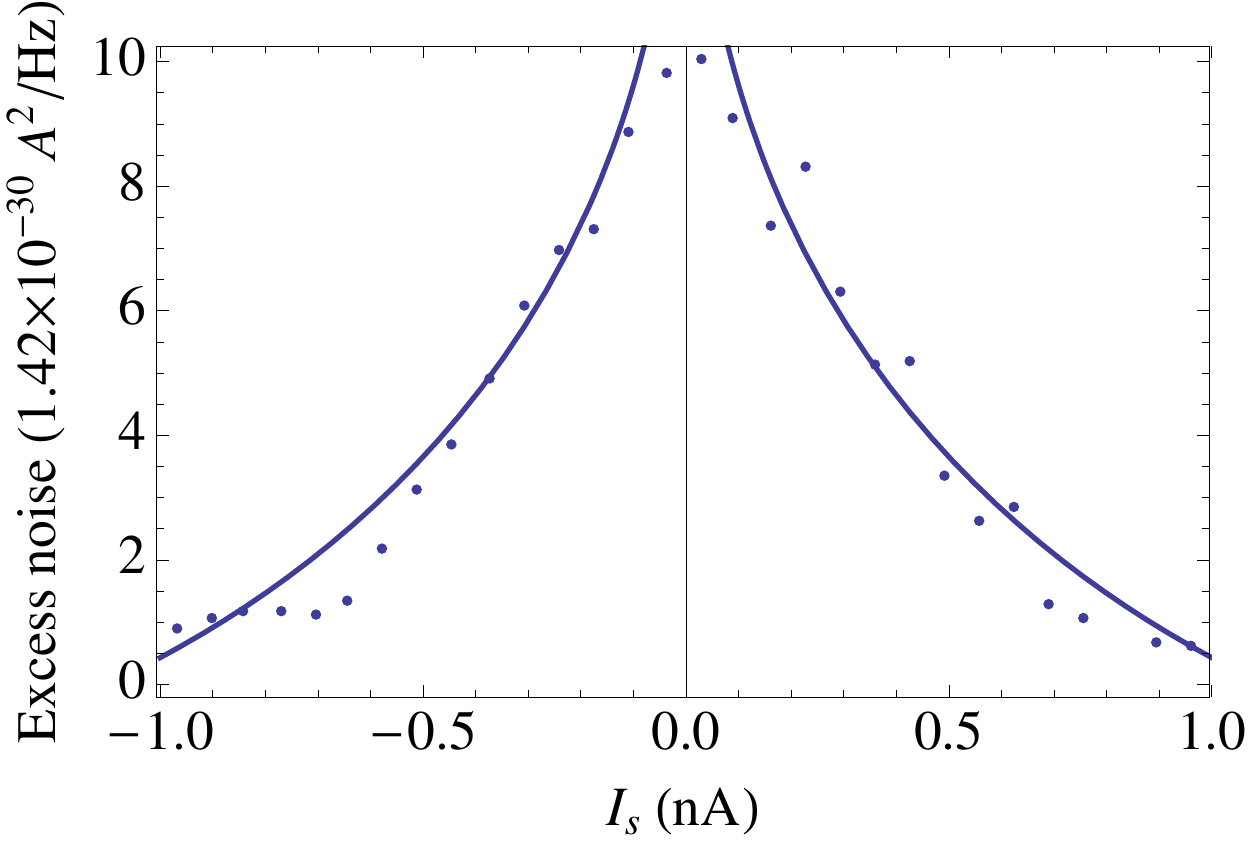}
        \caption{\textbf{Excess shot noise as a function of source current at finite $I_n$ at $\nu = 2/3$.} Dots show the measured shot noise at the indicated value of $I_n$, where the contribution at $I_n=0$ has been subtracted. Lines indicate the theoretical prediction.}\label{fig:BidDataFig3a}
\end{figure*}

In Fig~\ref{fig:BidDataFig3a} we show the excess shot noise $\Delta S_{tun}$ as a function of the source current $I_s$, for several values of injected current, $I_n$. The dots show the experimental data from Fig 3a in Ref~\onlinecite{Bid10}, where the noise at $I_n=0$ has been subtracted from the noise measured at finite $I_n$. The lines show our theoretical prediction, described in Sec~\ref{sec:weak-tunneling-bosons} and shown explicitly in Appendix~\ref{sec:compute23}  for the $\nu = 2/3$ edge:
\begin{align}
\Delta &S_{tun} \! = 2a \left( \frac{1}{3}\right)^2 \!\! \text{sgn}(I_n)
\left( \text{sgn}(c\sqrt{|I_n|} -\pi I_s )|\pi I_s -c\sqrt{|I_n|} |^{4/3}    \right. \nonumber\\
& \left. + \text{sgn}(\pi I_s +c\sqrt{|I_n|} )|\pi I_s +c\sqrt{|I_n|}  |^{4/3}-\frac{8}{3}c\sqrt{|I_n|}  |\pi I_s |^{1/3} \right) \label{eq:excess-noise-symm}
\end{align}
where $a \propto |\lambda_1|^2|\Lambda_1|^2$ and $c = \text{sgn}(I_n) |\Lambda_1|^{-1}/\sqrt{2\pi} $ . We have taken $ |\Lambda_1|=|\Lambda_2|$, consistent with the symmetry of the measured data under $I_s \rightarrow -I_s$. From here on, we will take the tunneling amplitudes to be constant for simplicity.\footnote{Experimental data on transmission, as in Ref~\onlinecite{Bid10}, is inconsistent with this assumption, but it does not qualitatively affect our point.} The constant $a$ is only known up to proportionality, as described in Appendix~\ref{sec:boson-calc}. The factor of 2 in front is consistent with the definition of shot noise in Ref~\onlinecite{Bid10}. Fitting $\Delta S_{tun}$ at $I_n = -.5, -1.0, -1.5,$ and $-2.0$ nA and averaging the best fit values yields $a = 4.4\times 10^{-30} \rm{A}^2 / \rm{Hz }/ \rm{nA}^{4/3}$ and $c=\text{sgn}(I_n)4.9$ nA$^{1/2}$. Fig~\ref{fig:BidDataFig3a} shows the excess noise at all four values of $I_n$ overlaid with the theoretical prediction using the averaged best fit values. The fits show that a change in temperature is not necessary to fit the excess noise that is measured. 

\begin{figure}[t]\centering
	\includegraphics[width=.48\textwidth]{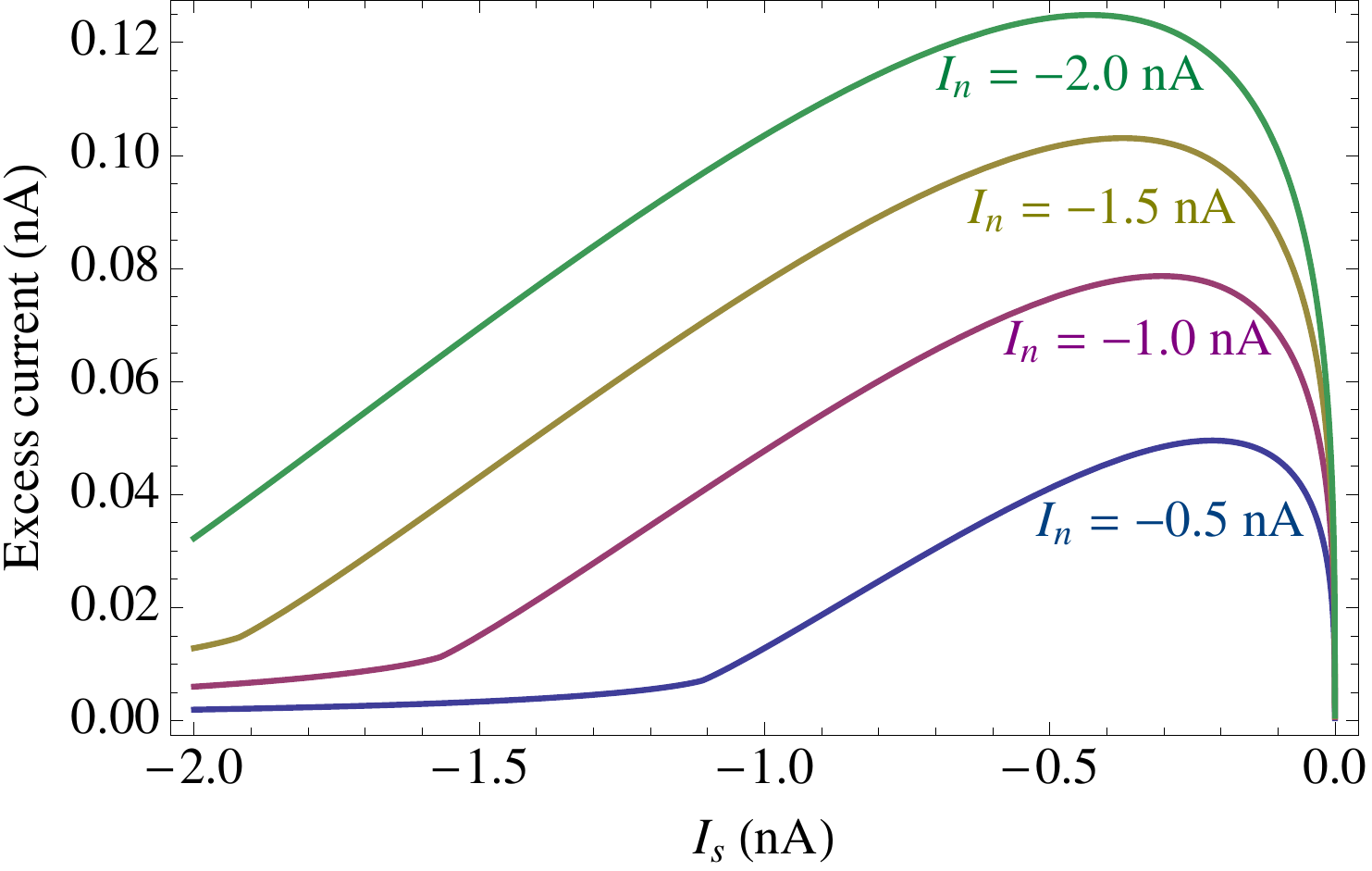}
	\caption{\textbf{Prediction of excess current when the injection lead is weakly coupled to the edge of the Hall bar at $\nu = 2/3$.} The absolute value in Eq~(\ref{eq:excess-current-symm}) causes the kink in each curve, which would be smooth at finite temperature.}
	\label{fig:ExcessCurrent}
\end{figure}

We then use the fit parameters to predict the change in current, $\Delta I_{tun}$, that results from injecting the neutral current, as described in Sec~\ref{sec:weak-tunneling-bosons} and computed in Appendix~\ref{sec:compute23}:
\begin{align}
\Delta I_{tun} &= \frac{a}{e} \frac{1}{3} \text{sgn}(I_n )\left(-|\pi I_s -c\sqrt{|I_n|} |^{4/3} + |\pi I_s + c\sqrt{|I_n|} |^{4/3} \right.\nonumber\\
& \left. -\frac{8}{3}c\sqrt{|I_n|}\text{sgn}(I_s)|\pi I_s|^{1/3}\right)  
\label{eq:excess-current-symm}
\end{align}
This prediction for excess current is shown in Fig~\ref{fig:ExcessCurrent} using the best-fit values. $\Delta I_{tun}$ has the opposite sign as $I_s$ and a maximum magnitude of .12nA when $I_n = -2$nA. Given that $I_n$ is measured in tenths of nanoamps, we expect $\Delta I_{tun} = .12$nA to be observable. This prediction might explain the slight decrease in transmission in Fig 3a of Ref~\onlinecite{Bid10}, but it is difficult to discern from the measurement whether the effect is real. It would helpful to increase $I_n$ further and observe whether the change in transmission becomes significant.

\subsection{Theoretical prediction of excess current for the $\nu = 2/3$ system with a temperature difference between the edges}

\begin{figure}[h]\centering
	\includegraphics[width=.48\textwidth]{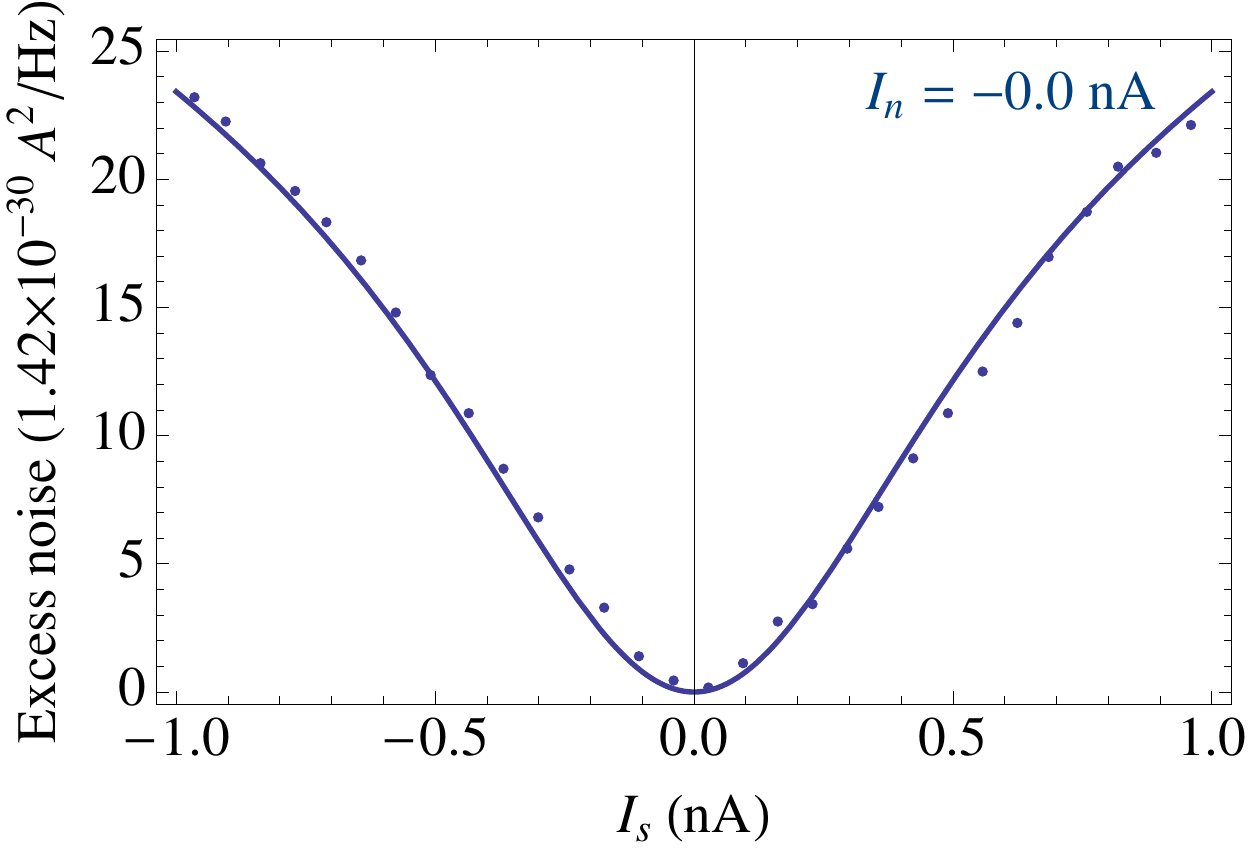}
	\caption{\textbf{Excess noise when $I_n=0$ at $\nu = 2/3$}. Dots show the measured noise while lines show the theoretical best fit.}
	\label{fig:excess-noise-strong}
\end{figure}

\begin{figure}[h]\centering
	\includegraphics[width=.48\textwidth]{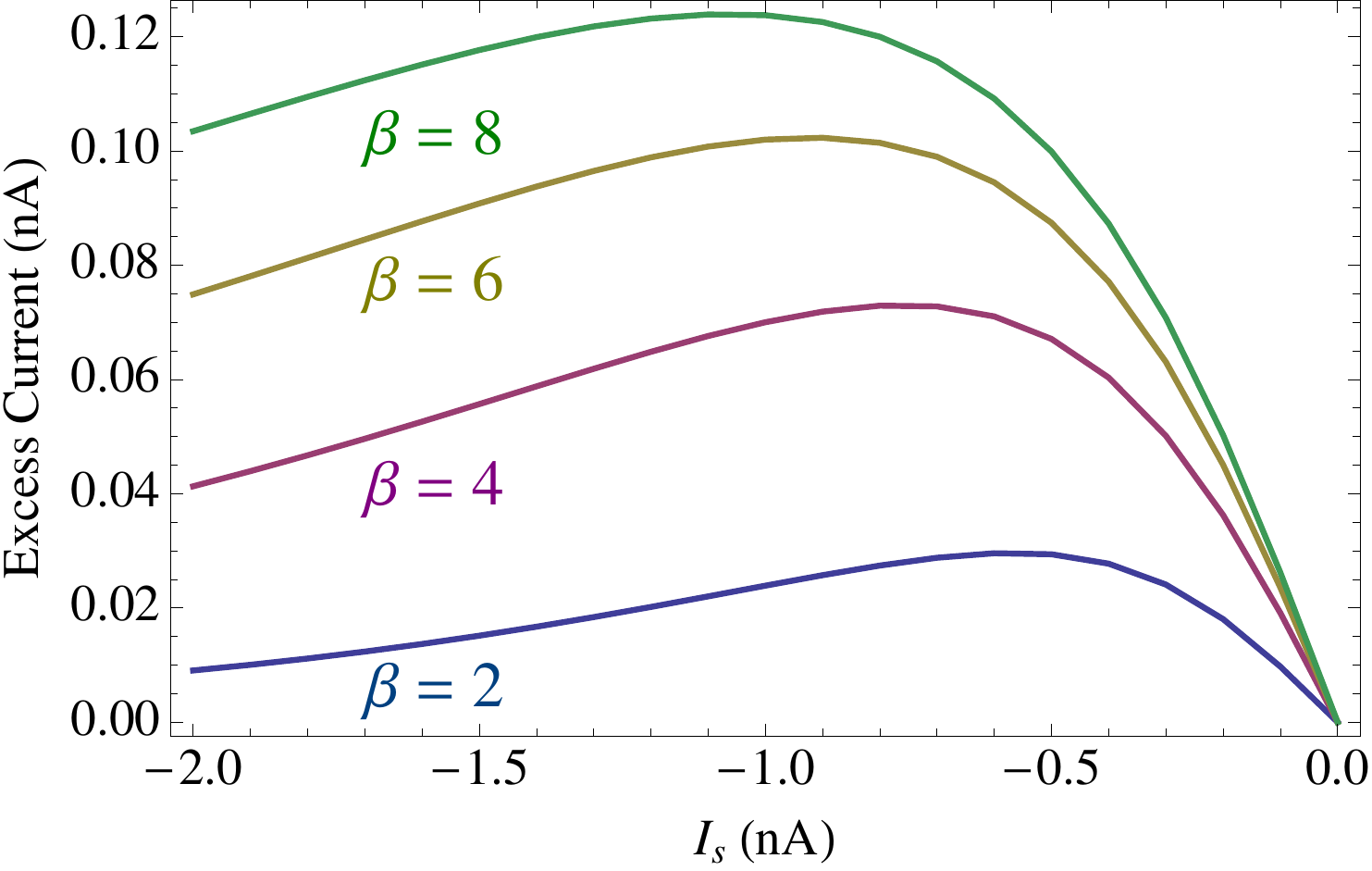}
	\caption{\textbf{Prediction of excess current in the temperature difference model at $\nu = 2/3$.}}
	\label{fig:strong-tunneling-excess-current}
\end{figure}

In the temperature difference model, the excess noise at $I_n = 0$ is given by Eq~(\ref{eq:excess-noise-strong}) applied to the $\nu = 2/3$ edge,
\begin{align} &S_{tun}|_{T_T = T_B} = -\frac{4}{\pi}(2\pi T_B)^{1/3} |\lambda_1|^2 \nonumber\\
& \times  \left[ \left( \frac{1}{3}\right)^2\sinh (\frac{\pi \alpha}{2}) B\left(\frac{2}{3}+i\frac{\alpha}{2}, \frac{2}{3}-i\frac{\alpha}{2} \right) i\psi \left(\frac{2}{3}+i\frac{\alpha}{2} \right)  \right. \nonumber\\
&+\left. \theta \left( \frac{2}{3} \right)^2 \sinh(\pi\alpha)B\left( \frac{2}{3} +i\alpha, \frac{2}{3}-i\alpha\right)i\psi\left( \frac{2}{3} + i\alpha \right) \right] + h.c.
\end{align}
where $ \alpha = I_s/T_B$ and $\theta = |\lambda_2|^2/|\lambda_1|^2 $. There is an extra factor of 2, consistent with the definition of shot noise used in Ref~\onlinecite{Bid10}. Using the best-fit value of $\theta = .39$ obtained in Ref~\onlinecite{Shtanko14}, we fit the shot noise at $I_n=0$ to find $T_B = 48$mK and $|\lambda_1|^2  = 1.8\times 10^{-29}{\rm K}^{-1/3}{\rm A}^2{\rm Hz}^{-1}$. These values yield the fit in Fig~\ref{fig:excess-noise-strong} and allow us to predict the magnitude of excess current using Eqs~(\ref{eq:tunn-current-strong}) and (\ref{eq:strong-integral}):
\begin{align}
\Delta I_{tun}&= \text{sgn}(\omega_0) 4C \left(  \int_0^\infty dx  \frac{\beta^{2/3}\left(\frac{1}{3}\sin(\alpha x)+\frac{2}{3}\theta \sin(2\alpha x) \right)}{(\sinh(\beta x))^{2/3}(\sinh(x))^{2/3}}\right. \nonumber\\
& \left. -\frac{ \frac{1}{3} \sin(\alpha x)+\frac{2}{3}\theta \sin(2\alpha x) }{(\sinh(x))^{4/3}}\right) \label{eq:delta-I-tun-strong-23}
\end{align}
where $C=\left( \pi T_B\right)^{1/3}|\lambda_1|^2 \sin(2\pi/3) $ and $\beta = T_T/T_B$. For simplicity we have taken the tunneling amplitudes to be constant. In Ref~\onlinecite{Shtanko14}, the authors find the fit $\beta = 1+5.05 | I_n \text{ nA}^{-1}| ^{.54} $, which predicts that $\beta$ increases from 1 to 8 as $I_n$ is turned up to 2 nA. The predicted current is shown in Fig~\ref{fig:strong-tunneling-excess-current}, where the maximum change in current is seen to be .12 nA. Coincidentally, this is the same magnitude as predicted from the weak tunneling model. We believe this current to be observable in experiment.

The theoretical prediction is a good fit to the data but both the best-fit temperature of $T_B = 48$mK and the best-fit temperature increase by a factor of $\beta \sim 8$ at $I_n = -2$ nA are significantly larger than the increase from 10mK to 25mK estimated in Ref~\onlinecite{Bid10}. This might be attributed to a discrepancy between the modified free-fermion model used in Ref~\onlinecite{Bid10} to fit the data and the Luttinger liquid model used here. An independent measurement in Ref~\onlinecite{Gurman12} found the temperature of a $\nu = 2/3$ edge to increase from 30mK to 130mK over a similar range of $I_n$, using quantum dot thermometry.\cite{Venkatachalam12,Viola12} However, it is not clear whether the measurement in Ref~\onlinecite{Gross12} of excess noise that varies with transmission probability at a QPC that has neutral modes impinging from both edges can be explained completely by the temperature difference model: since both edges would be raised to the same temperature there would no longer be a temperature gradient across the QPC.

\section{Conclusions}
We have used a theoretical model to predict that for an edge with counter-propagating neutral modes, current injected downstream of the QPC causes a change in both the shot noise and tunneling current across the QPC. In the specific case of $\nu = 2/3$, we have compared our expression for excess shot noise to the values measured in experiment to determine two fitting parameters. We then used these fitting parameters to predict that the magnitude of the tunneling current could decrease by as much as .1 nA when downstream current is injected. This current should be barely large enough to be observed experimentally; it would increase if the injected current increased, which could perhaps be a subject of future work.

In a different model of the system which incorporates the injected current as a temperature difference between the two edges of the Hall bar, fitting the experimental shot noise data also (coincidentally) predicts that the tunneling current should change by approximately .1 nA. This fit yields a ratio of the temperatures ${T_T}/{T_B} \approx 8$, which seems high, but is not unreasonable, given the measurements in Ref~\onlinecite{Gurman12}.

It is likely that the physical edge is described by a theory that contains elements from both models. Since both models predict a measurable change in transmission when current is injected downstream of the QPC, we believe this change is a real feature. The prediction runs counter to the intuition that motivated the seminal experiment in Ref~\onlinecite{Bid10}; hence, it would be interesting to systematically study this effect experimentally. We expect that our theory could be applied to the $\nu = 5/2$ state by matching the scaling dimensions of the tunneling operators, but the non-Abelian nature of some candidate states might prove to be non-trivial. Another future direction would be to study the dependence of the shot noise and tunneling current on the distance between the current injection site and the QPC at $x=0$.

\section{Acknowledgments}
We would like to thank Vadim Cheianov, David Clarke, Moty Heiblum, Roman Lutchyn, Michael Mulligan, Bernd Rosenow and Kyrylo Snizhko for helpful discussions.
We acknowledge the support of the DARPA Quest program.
J.C. acknowledges the support of the National Science Foundation Graduate Research Fellowship under Grant No. DGE1144085.

\appendix
\section{Non-equilibrium correlation functions}
\label{sec:Keldysh}

Because current flows from one edge of the Hall bar to the other, the system at hand is not in equilibrium. Hence, we rely on the Schwinger-Keldysh formalism\cite{Schwinger61,Keldysh65} to calculate correlators: to do this, we double the time contour so that $t$ goes from $-\infty$ to $+\infty$ and back to $-\infty$ again. It is convenient to label the forward and backward moving time paths by a superscript $\mu = \pm$, so that each time $t$ is now labeled by $t^\mu$ and an integral over all times is now $\int_K dt \equiv \int_{-\infty}^{\infty}dt^+ + \int_\infty^{-\infty}dt^- = \int_{-\infty}^{\infty}dt^+ - \int_{-\infty}^{\infty}dt^-$. Correlation functions now depend on which side of the contour the times lie on. A thorough explanation of bosonic correlators is given in Refs~\onlinecite{Chamon95} and \onlinecite{Chamon96}. Here we state the result for right- or left-moving bosonic fields $\phi_{R/L}$:
\begin{align} \langle \phi_{R/L}& (t_1^\mu,x_1)\phi_{R/L}(t_2^\nu,x_2) \rangle \equiv -\ln G^{\mu\nu}_{R/L}(t_1-t_2,x_1-x_2)    \nonumber\\
&= -\ln \left(\epsilon + iK^{\mu\nu}(t_1-t_2)(t_1-t_2 \mp (x_1-x_2)) \right) \end{align}
where 
\begin{equation} K^{\mu\nu}(t)=\Theta (\mu\nu)\text{sgn}(\mu t) + \Theta (-\mu\nu)\text{sgn}\nu \end{equation} 
and $\epsilon$ is a small positive number.
Fermions get an extra minus sign:
\begin{equation} \langle \psi_{R/L}(t_1^\mu,x_1)\psi_{R/L}^\dagger(t_2^\nu,x_2) \rangle = \frac{K^{\mu\nu}(t_1-t_2)}{G^{\mu\nu}_{R/L}(t_1-t_2,x_1-x_2) }\end{equation}

Similarly to the minus sign for fermions, we need to consider the Klein factors for the bosonic tunneling operators when we consider systems with two QPCs, as in Sec~\ref{sec:weak-tunneling}. Without the Klein factors, the tunneling terms in the Lagrangian at $x=0$ and $x=a$ generically do not commute:
\begin{align}&\left( e^{iq\phi_T(0)}e^{iq\phi_B(0)}\right)\left(e^{i\Phi(a)}e^{-ir\phi_T(a)}\right) \nonumber\\
&= \left(e^{i\Phi(a)}e^{-ir\phi_T(a)} \right)\left(e^{iq\phi_T(0)}e^{iq\phi_B(0)}\right) e^{i\pi (q_nr_n-q_cr_c)\text{sgn}(a)}\end{align}
(We have bosonized the external lead $\Psi \sim e^{-i\Phi}$ and shown the spatial arguments but suppressed the time arguments.) However, because the product of the pair of operators in each set of parenthesis is bosonic, they are physically required to commute. This discrepancy is resolved by including Klein factors. The prescription is as follows: to each tunneling quasiparticle or electron operator, we attach a Klein factor $\kappa_{x,T/B/E}$, where the subscript $x=0,a$ indicates the position at which the tunneling operator acts and $T/B/E$ indicates the edge: top/bottom/external lead. For example: $e^{ iq\phi_{T}(0)}\rightarrow \kappa_{0,T}^\dagger e^{ iq\phi_{T}(0)}$. 
The $\kappa$s then must satisfy the commutation relation:
\begin{equation} \kappa_{0,T}^\dagger\kappa_{0,B}\kappa_{a,E}^\dagger\kappa_{a,T} = \kappa_{a,E}^\dagger\kappa_{a,T}  \kappa_{0,T}^\dagger\kappa_{0,B} e^{-i\pi(q_nr_n-q_cr_c)\text{sgn}(a)}\end{equation}
Notice that in the calculation of any physical quantity, the $\kappa$s will come in pairs $\kappa_{0/a,T}^\dagger\kappa_{0/a,B/E}$. Following Ref~\onlinecite{Guyon02}, it is convenient to bosonize these pairs:
\begin{equation} \kappa_{0,T}^\dagger \kappa_{0,B} = e^{-i\theta_0}, \quad \kappa_{a,E}^\dagger\kappa_{a,T} = e^{-i\theta_a} \end{equation}
Using $\text{sgn}(a)=-1$, we find $[\theta_0,\theta_a]= -i\pi(q_nr_n-q_cr_c)$. If $q_cr_c>q_nr_n$ then conventional raising and lowering operators can be defined by $a=\frac{1}{\sqrt{2\pi(q_cr_c-q_nr_n)}}\left(\theta_0+i\theta_a\right)$, where $\langle a^\dagger a\rangle_0 = 0, \langle aa^\dagger\rangle = 1$. This readily yields $\langle \theta_0\theta_a\rangle = -\langle \theta_a\theta_0\rangle= i\langle\theta_0\theta_0\rangle = i\langle \theta_a\theta_a\rangle = i\frac{\pi}{2}\left( q_cr_c-q_nr_n\right)$. (If $q_nr_n>q_cr_c$, the roles of $a$ and $a^\dagger$ are reversed). Finally, since the system is not in equilibrium, we will actually need to use $\langle \theta_0(t_1^\mu)\theta_a(t_2^\nu) \rangle= i\frac{\pi}{2}K^{\mu\nu}(t_1-t_2)(q_cr_c-q_nr_n)$.

The Klein factors drop out of the leading order current and noise calculations, which only include one QPC, but are important in computing the excess noise and current.

\section{Fermionic excess current and noise calculations}
\label{sec:fermion-calc}
Here we show the details of how to find $\Delta I_{tun}$ to leading order as written in Eq~(\ref{eq:define-corr-function}) using the correlators described in Sec~\ref{sec:Keldysh}. As a warm-up, we calculate the leading order noise and current across a QPC using the Keldysh formalism. Model the current injection at $x=a$ by the tunneling term 
\begin{equation} \mathcal{L}_{tun} = -\lambda e^{-i\omega_0 t} \psi_1^\dagger \cdots \psi_n^\dagger \psi_{n+1} \cdots \psi_{2n}\delta(x-a)  + h.c.\label{eq:tun-fermions-general}\end{equation}
where each of the fermion fields obeys the Lagrangian (\ref{eq:free-fermion}) for some chirality that will not be important here. The current operator corresponding to this tunneling term is $I_{tun} = e \frac{d}{dt}N_T = ie[N_T,H] = -ier_c\lambda e^{-i\omega_0 t} \psi_1^\dagger \cdots \psi_n^\dagger \psi_{n+1} \cdots \psi_{2n}  + h.c.$, where $N_T$ is the number operator for the charged electron fields on the top edge and $r_c$ is an integer that counts the charged fields. Then the current across the QPC is given to leading order by
\begin{align} &\langle I_{tun} \rangle_0 = \langle I_{tun} i\int dt\mathcal{L}_{tun} \rangle\nonumber\\
&= er_c |\lambda|^2 \!\! \int_K \!\! dt \prod_{j=1}^{n} \langle \psi_j^\dagger(0^+) \psi_j(t^\mu) \rangle \!\!\!\!  \prod_{j=n+1}^{2n}\!\! \langle \psi_j(0^+) \psi_j^\dagger(t^\mu) \rangle e^{-i\omega_0 t} + h.c. \nonumber\\
&= er_c|\lambda|^2 \int_K \left( \frac{K^{+\mu}(-t)}{G^{+\mu}(-t)}\right)^{2n}(-2i\sin(\omega_0 t)) \nonumber\\
&= -2ier_c |\lambda|^2 \left( \int_{-\infty}^\infty dt \frac{\left(\text{sgn}(-t)\right)^{2n}}{(\epsilon+i|t|)^{2n}}- \frac{(-1)^{2n}}{\left( \epsilon+it\right)^{2n}}  \right) \sin(\omega_0 t)\nonumber\\
&= \frac{2\pi}{\Gamma(2n)}er_c|\lambda|^2 (\omega_0)^{2n-1}
\label{eq:I-tun-0-Keldysh}
\end{align}
We have placed the time $0$ on the $+$ side of the Keldysh contour (as long as it is fixed, either side is correct) and integrated the time $t$ over both sides. We have suppressed the space index because it does not enter. Notice that in the second to last line, the integrand disappears when $t>0$, which enforces chirality. This result yields Eqs~(\ref{eq:I-tun-0}) and (\ref{eq:I-inj-0}).

The noise across the junction $S(t) = \frac{1}{2}\lbrace I_{tun}(t^+),I_{tun}(0^-) \rbrace$ is similarly computed and its Fourier transform $S(\omega)$ is found to be 
\begin{equation} S(\omega) = \frac{1}{2}er_c I_{tun} \left( |1- \omega/\omega_0|^{2n-1}+|1+\omega/\omega_0|^{2n-1} \right)
\end{equation}
which yields the usual relation in the zero-frequency limit $S(\omega=0)=er_cI_{tun}$. However, when there are multiple tunneling terms $\mathcal{L}_{tun,k}$ that tunnel different amounts of charge $n_k$, the proportionality of the total current and noise that are measured is lost, and $S(\omega=0)/I_{tun} = e \left( \sum_k n_k I_{tun,k} \right)/ \left( \sum_k  I_{tun,k} \right)$.

\subsection{Excess current}
\label{sec:fermion-excess-current}

We now tackle the next order in perturbation theory to find $\Delta I_{tun}$. Using Eqs~(\ref{eq:L2-tun}), (\ref{eq:L2-inj}), and (\ref{eq:define-corr-function}) and the correlation functions of the previous section yields
\begin{align}
&\frac{\Delta I_{tun}}{e|\lambda_2|^2|\Lambda_2|^2} \!=\!  2\!\! \int_K\! dt_1^\mu dt_2^\nu dt_3^\sigma \frac{2i\sin(\omega_0 t_1 +\omega_1(t_3-t_2))}{\left(G^{+\mu}(-t_1,0)\right)^2 \!\!  \left(G^{\nu\sigma}(t_2-t_3,0)\right)^2} \nonumber\\
&\! \times \!\! \Big( \! F_L(0,1,2,3)+F_L(0,3,1,2) \Big) \!\! \Big( \! F_R(0,1,2,3)+F_R(0,3,1,2) \Big)
\label{eq:delta-I-tun-Keldysh}
\end{align}
where
\begin{align}
&F_{L/R}(i,j,k,l) = \frac{K^{\mu_i\mu_j}(t_i-t_j)K^{\mu_k\mu_l}(t_k-t_l)}{G^{\mu_i\mu_j}_{L/R}(t_i-t_j,x_i-x_j)G^{\mu_k\mu_l}_{L/R}(t_k-t_l,x_k-x_l)}
\end{align}
and $t_0=0, x_0=x_1=0, x_2=x_3=a$.
The subscripts $R/L$ have been omitted where there is no $x$ argument. The factor of 2 in front of Eq~(\ref{eq:delta-I-tun-Keldysh}) is from another set of terms that occurs in Eq~(\ref{eq:define-corr-function}) which is related by $t_2 \leftrightarrow t_3$.


We now consider each term in Eq~(\ref{eq:delta-I-tun-Keldysh}) when the product in the second line is expanded. The term $F_L(0,1,2,3)F_R(0,1,2,3)$ disappears when all parts of the Keldysh contour are summed over (even before integration) because it does not mix times 0 and $t_1$ with $t_2,t_3$.
We now consider the term $F_L(0,1,2,3)F_R(0,3,1,2)$, which, when each contribution of the Keldysh contour is added, contributes to the right hand side of Eq~(\ref{eq:delta-I-tun-Keldysh})
\begin{widetext}
\begin{align}
\Delta & I_{tun,12}=\int_{-\infty}^0 dt_1dt_2dt_3 \frac{4i\sin(\omega_0 t_1 +\omega_1(t_3-t_2))}{\left(\epsilon-it_1\right)^3} \left( \frac{1}{(\text{sgn}(t_2-t_3)\epsilon +i(t_2-t_3))^3(\epsilon-i(t_3-a))(\text{sgn}(t_1-t_2)\epsilon+i(t_1-t_2+a))} \right. \nonumber\\
& -\frac{1}{(-\epsilon +i(t_2-t_3))^3(-\epsilon-i(t_3-a))(\text{sgn}(t_1-t_2)\epsilon+i(t_1-t_2+a))} 
-\frac{1}{(\epsilon +i(t_2-t_3))^3(\epsilon-i(t_3-a))(-\epsilon+i(t_1-t_2+a))} \nonumber\\
&\left. +\frac{1}{(-\text{sgn}(t_2-t_3)\epsilon +i(t_2-t_3))^3(-\epsilon-i(t_3-a))(-\epsilon+i(t_1-t_2+a))} \right) + h.c. \nonumber\\
&= \Theta(-a) \pi \int_{-\infty}^{\infty} dt_1dt_3 4i\sin(-\omega_0 t_1 + \omega_1 t_3)\left( \frac{1}{(\epsilon-it_1)^3}+ \frac{1}{(\epsilon+it_1)^3} \right)\left( \frac{1}{(\epsilon-it_3)^3(\epsilon-i(t_3-t_1))}- \frac{1}{(\epsilon+it_3)^3(\epsilon+i(t_3-t_1))}\right) \nonumber\\
%
&=8\pi^2 \Theta(-a) \int_{-\infty}^\infty dt_1 \left( \frac{1}{(\epsilon-it_1)^3}+ \frac{1}{(\epsilon+it_1)^3} \right)\left( \frac{-\omega_1t_1 \cos(\omega_0 t_1)+ \sin(\omega_0 t_1)(1-\frac{1}{2}t_1^2\omega_1^2)-\sin((\omega_0-\omega_1)t_1) }{t_1^3} \right)
\end{align}
\end{widetext}
In the first line, the integral is only written for $t_i<0$ because the $t_i>0$ terms cancel; this is an example of how the Keldysh contour enforces chirality. To get the first equality, we used the expression for a delta-function
\begin{equation}
\lim_{\epsilon\rightarrow 0} \frac{\epsilon}{\epsilon^2 + x^2} = \pi \delta(x) \label{eq:delta-function}
\end{equation}
with some algebraic manipulations. The second equality is from doing the $t_3$ integral exactly.
The denominator $1/t_1^3$ in the first line should not be a concern because the limit of the entire term in parenthesis is a constant as $t_1 \rightarrow 0$. Consequently, we can push the pole at $t_1=0$ to either side of the imaginary axis in order to do the contour integral. This strategy yields
\begin{equation} \Delta I_{tun,12} =- \frac{2}{15} \pi^3 
\left( 10\omega_0^2\omega_1^3-5\omega_0\omega_1^4+\omega_1^5 \right)\Theta(-a) \label{eq:tunn-current-fermions-appendix-0}
\end{equation}
We now impose the condition that the neutral fermions are Majorana fermions. This generates an extra term in the 4-point correlation function $\langle \psi_0(0,0)\psi_0(t_1,0)\psi_0(t_2,a)\psi_0(t_3,a)\rangle$ of Eq~(\ref{eq:delta-I-tun-Keldysh}) that is equivalent to taking $\omega_1 \rightarrow -\omega_1$. Consequently, when the neutral fermions are Majorana fermions, 
\begin{equation} \Delta I_{tun,12} = \frac{4}{3} \pi^3 
\omega_0\omega_1^4\Theta(-a) \label{eq:tunn-current-fermions-appendix}
\end{equation}
This yields Eq~(\ref{eq:tunn-current-fermions}) in the main text.

The next term in Eq~(\ref{eq:delta-I-tun-Keldysh}) to consider is the term that contains the product $F_L(0,3,1,2)F_R(0,1,2,3)$. By symmetry, this term will yield Eq~(\ref{eq:tunn-current-fermions-appendix}) with $a\leftrightarrow -a$. We are now left with only one more term in Eq~(\ref{eq:delta-I-tun-Keldysh}), that which contains $F_L(0,3,1,2)F_R(0,3,1,2)$ and contributes to the right hand side of Eq~(\ref{eq:delta-I-tun-Keldysh})
\begin{widetext}
\begin{align}
\Delta & I_{tun,22}=\int_{-\infty}^0 dt_1dt_2dt_3 \frac{4i\sin(\omega_0 t_1 +\omega_1(t_3-t_2))}{\left(\epsilon-it_1\right)^2}\nonumber\\
&\times \left( \frac{1}{(\text{sgn}(t_2-t_3)\epsilon +i(t_2-t_3))^2(\epsilon-i(t_3-a))(\text{sgn}(t_1-t_2)\epsilon+i(t_1-t_2+a))(\epsilon-i(t_3+a))(\text{sgn}(t_1-t_2)\epsilon+i(t_1-t_2-a))} \right. \nonumber\\
& -\frac{1}{(-\epsilon +i(t_2-t_3))^2(-\epsilon-i(t_3-a))(\text{sgn}(t_1-t_2)\epsilon+i(t_1-t_2+a))(-\epsilon-i(t_3+a))(\text{sgn}(t_1-t_2)\epsilon+i(t_1-t_2-a))} \nonumber\\
&-\frac{1}{(\epsilon +i(t_2-t_3))^2(\epsilon-i(t_3-a))(-\epsilon+i(t_1-t_2+a))(\epsilon-i(t_3+a))(-\epsilon+i(t_1-t_2-a))} \nonumber\\
&\left. +\frac{1}{(-\text{sgn}(t_2-t_3)\epsilon +i(t_2-t_3))^2(-\epsilon-i(t_3-a))(-\epsilon+i(t_1-t_2+a))(-\epsilon-i(t_3+a))(-\epsilon+i(t_1-t_2-a))} \right) + h.c. \nonumber\\
%
&=\int_{-\infty}^0 dt_1dt_2dt_3 \frac{4i\sin(\omega_0 t_1 +\omega_1(t_3-t_2))}{\left(\epsilon-it_1\right)^2}\left( \frac{1}{-4a^2} \right) \nonumber\\
&\times \left( \frac{1}{(\text{sgn}(t_2-t_3)\epsilon +i(t_2-t_3))^2(\epsilon-i(t_3-a))(\text{sgn}(t_1-t_2)\epsilon+i(t_1-t_2+a))} \right. \nonumber\\
& -\frac{1}{(-\epsilon +i(t_2-t_3))^2(-\epsilon-i(t_3-a))(\text{sgn}(t_1-t_2)\epsilon+i(t_1-t_2+a))}-\frac{1}{(\epsilon +i(t_2-t_3))^2(\epsilon-i(t_3-a))(-\epsilon+i(t_1-t_2+a))} \nonumber\\
&\left. +\frac{1}{(-\text{sgn}(t_2-t_3)\epsilon +i(t_2-t_3))^2(-\epsilon-i(t_3-a))(-\epsilon+i(t_1-t_2+a))} \right) + h.c. \nonumber\\
&=-\frac{\pi}{a^2}\int_{-\infty}^\infty dt_1dt_2 i\sin(\omega_0t_1-\omega_1 t_2 )\left( \frac{1}{(\epsilon-it_1)^2}- \frac{1}{(\epsilon+it_1)^2} \right)\left( \frac{1}{(\epsilon-it_2)^2(\epsilon-i(t_2+t_1))}+\frac{1}{(\epsilon+it_2)^2(\epsilon+i(t_2+t_1))} \right) \nonumber\\
&= -\frac{2\pi^2}{a^2}\int_{-\infty}^\infty dt_2 \left( \omega_0\cos(\omega_1 t_2)\left( \frac{1}{(\epsilon-it_2)^3}+\frac{1}{(\epsilon+it_2)^3} \right) -\sin(\omega_1t_2) \left( \frac{i}{(\epsilon-it_2)^4} - \frac{i}{(\epsilon+it_2)^4} \right)\right) \nonumber\\
&= -\frac{2\pi^3}{a^2} \left( \omega_0\omega_1^2  +\frac{1}{3} \omega_1^3\right)
\end{align}
\end{widetext}
where in the first equality, we have taken $a<0$ and removed the infinitesimal $\epsilon$'s from terms in the denominator that are not small and approximated their values (specifically, the integral is dominated by $t_3 ,t_2\approx a<0$ and $t_1\approx 0$, so $\pm \epsilon-i(t_3+a) \approx -2ia$ while $\pm \epsilon+i(t_1-t_2-a) \approx -2ia$. By symmetry, we will get the same answer when $a>0$. The second equality again utilizes the delta-function identity (\ref{eq:delta-function}), along with some algebra. The third equality uses the derivative of the delta-function identity and integration by parts. In the main text, we consider the case where the injected current is far away from the tunneling QPC, so that $|a| \gg 1/\omega_0,1/\omega_1$, and hence $\Delta I_{tun,22} \ll \Delta I_{tun,12}$. However, $\Delta I_{tun,22}$ is present, and in a future experiment where $a$ is of the same scale as the length scale set by the voltages, we would expect it to have an effect, which is independent of the sign of $a$.

\subsection{Excess noise}

The excess shot noise is defined by 
\begin{align}
\Delta &S_{tun}(t) = \frac{1}{2} \left(\langle \lbrace I_{2}(t),I_{2}(0) \rbrace \rangle |_{\Lambda_i} -  \langle \lbrace I_{2}(t),I_{2}(0) \rbrace \rangle |_{\Lambda_i =0}\right) \nonumber\\
&=\frac{1}{2} \langle  \lbrace I_{2}(t),I_{2}(0) \rbrace \left(i\int dt_2 \mathcal{L}_{inj}^{2}\right)\left( i\int dt_3\mathcal{L}_{inj}^{2} \right) \rangle_0
\end{align}
We will compute $\Delta S_{tun}(\omega) = \int dt e^{i\omega t}\Delta S_{tun}(t)$ using the tunneling terms Eqs~(\ref{eq:L2-tun}) and (\ref{eq:L2-inj}):
\begin{align}
&\frac{\Delta S_{tun}(\omega)}{e^2|\lambda_{2}|^2|\Lambda|^2} = - \int_{-\infty}^\infty dt 2\cos(\omega t)\nonumber\\
&\times \int_K dt_2^\nu dt_3^\sigma \frac{2\cos(\omega_0 t + \omega_1(t_3-t_2))}{(G^{+-}(-t,0))^2(G^{\nu\sigma}(t_2-t_3,0))^2} \nonumber\\
&\times \left( F_L(0,1,2,3)+F_L(0,3,1,2)\right)\left( F_R(0,1,2,3)+F_R(0,3,1,2) \right) \label{eq:delta-S-tun-Keldysh}
\end{align}
The factor of $\frac{1}{2}$ that is in the definition of $S(t)$ is cancelled by a factor of 2 that comes from a different term that takes $t_2 \leftrightarrow t_3$. Eq~(\ref{eq:delta-S-tun-Keldysh}) looks very similar to Eq~(\ref{eq:delta-I-tun-Keldysh}) so we can provide an abbreviated analysis.

The term that contains the product $F_L(0,1,2,3)F_R(0,1,2,3)$ disappears, as in the excess current calculation. We now consider the term that contains $F_L(0,1,2,3)F_R(0,3,1,2)$, the analogue of $\Delta I_{tun,12}$ defined in the previous section:
\begin{widetext}
\begin{align}
\Delta & S_{tun,12}(\omega) =4\int_{-\infty}^\infty \! \! \! \! dt \cos(\omega t) \int_{-\infty}^{0} \! \! \! \! dt_2dt_3 \frac{\cos(\omega_0 t +\omega_1(t_3-t_2))}{\left(\epsilon+it\right)^3} \left( \frac{1}{(\text{sgn}(t_2-t_3)\epsilon +i(t_2-t_3))^3(\epsilon-i(t_3-a))(\epsilon+i(t-t_2+a))} \right. \nonumber\\
& -\frac{1}{(-\epsilon +i(t_2-t_3))^3(-\epsilon-i(t_3-a))(\epsilon+i(t-t_2+a))} 
-\frac{1}{(\epsilon +i(t_2-t_3))^3(\epsilon-i(t_3-a))(-\epsilon+i(t-t_2+a))} \nonumber\\
&\left. +\frac{1}{(-\text{sgn}(t_2-t_3)\epsilon +i(t_2-t_3))^3(-\epsilon-i(t_3-a))(-\epsilon+i(t-t_2+a))} \right) \nonumber\\
&= \Theta(-a) 4\pi \int_{-\infty}^{\infty} \! \! \! \! dt dt_3 \cos(\omega t) \cos(\omega_0 t - \omega_1 t_3)\left( \frac{1}{(\epsilon+it)^3}- \frac{1}{(\epsilon-it)^3} \right)\left( \frac{1}{(\epsilon+it_3)^3(\epsilon-i(t-t_3))}- \frac{1}{(\epsilon-it_3)^3(\epsilon+i(t-t_3))}\right) \nonumber\\
&= \Theta(-a) 8\pi^2 i \int_{-\infty}^\infty \!\!\!\! dt \cos(\omega t) \left( \frac{1}{(\epsilon+it)^3}-\frac{1}{(\epsilon-it)^3} \right)\frac{1}{t^3}\left( \cos(\omega_0 t)(-1+\frac{1}{2}t^2\omega_1^2)-t\omega_1 \sin(\omega_0 t)+\cos((\omega_0 - \omega_1)t) \right)     \nonumber\\
&\xrightarrow{\omega =0}\Theta(-a)\frac{2\pi^3}{15}\left( -|\omega_0|^5 - 10\omega_1^2|\omega_0|^3+5\omega_1\text{sgn}(\omega_0)|\omega_0|^4 + |\omega_0 - \omega_1|^5 \right) \nonumber\\
&\xrightarrow{\omega=0} 
\begin{cases}  \Theta(-a)\frac{2\pi^3}{15}|\omega_1|^5 & |\omega_0| \ll |\omega_1| \\
-\Theta(-a)\frac{4\pi^3}{3}\omega_0^2|\omega_1|^3\text{sgn}(\omega_0\omega_1)  & |\omega_0| \gg |\omega_1|\end{cases}
\end{align}
\end{widetext}
To obtain the limits of integration in the first line, we used the fact that the integral is dominated by $t\approx 0$ and $t_2,t_3 \approx a$ because of the placement of the poles. The rest of the equalities follow similarly to the excess current calculation in the previous section. The result is that in either the limit $|\omega_0| \gg |\omega_1|$ or $|\omega_1| \gg |\omega_0|$, $|\Delta S_{tun,12}(\omega=0)| = e|\Delta I_{tun,12}|$, but generically, this proportionality does not hold. If the neutral fermions are Majorana fermions then the extra term in the 4-point correlation function contributes an overall factor of two.

We now consider the remaining terms in Eq~(\ref{eq:delta-S-tun-Keldysh}). The term containing $F_L(0,3,1,2)F_R(0,1,2,3)$ is the same as $\Delta S_{tun,12}$ with $a \rightarrow -a$. The final term is that containing $F_L(0,1,2,3)F_R(0,3,1,2)$. Similarly to the calculation of $\Delta I_{tun,22}$, we compute
\begin{widetext}
\begin{align}
\Delta & S_{tun,22}(\omega)= 4\int_{-\infty}^\infty \!\!\!\! dt \cos(\omega t) \int_{-\infty}^0 dt_2dt_3 \frac{\cos(\omega_0 t +\omega_1(t_3-t_2))}{\left(\epsilon+it\right)^2}\nonumber\\
&\times \left( \frac{1}{(\text{sgn}(t_2-t_3)\epsilon +i(t_2-t_3))^2(\epsilon-i(t_3-a))(\epsilon+i(t-t_2+a))(\epsilon-i(t_3+a))(\epsilon+i(t-t_2-a))} \right. \nonumber\\
& -\frac{1}{(-\epsilon +i(t_2-t_3))^2(-\epsilon-i(t_3-a))(\epsilon+i(t-t_2+a))(-\epsilon-i(t_3+a))(\epsilon+i(t-t_2-a))} \nonumber\\
&-\frac{1}{(\epsilon +i(t_2-t_3))^2(\epsilon-i(t_3-a))(-\epsilon+i(t-t_2+a))(\epsilon-i(t_3+a))(-\epsilon+i(t-t_2-a))} \nonumber\\
&\left. +\frac{1}{(-\text{sgn}(t_2-t_3)\epsilon +i(t_2-t_3))^2(-\epsilon-i(t_3-a))(-\epsilon+i(t-t_2+a))(-\epsilon-i(t_3+a))(-\epsilon+i(t-t_2-a))} \right) \nonumber\\
&=-\frac{1}{a^2}\int_{-\infty}^\infty \!\!\!\! dt \cos(\omega t) \int_{-\infty}^0 dt_2dt_3 \frac{\cos(\omega_0 t +\omega_1(t_3-t_2))}{\left(\epsilon+it\right)^2}  \left( \frac{1}{(\text{sgn}(t_2-t_3)\epsilon +i(t_2-t_3))^2(\epsilon-i(t_3-a))(\epsilon+i(t-t_2+a))} \right. \nonumber\\
& -\frac{1}{(-\epsilon +i(t_2-t_3))^2(-\epsilon-i(t_3-a))(\epsilon+i(t-t_2+a))}-\frac{1}{(\epsilon +i(t_2-t_3))^2(\epsilon-i(t_3-a))(-\epsilon+i(t-t_2+a))} \nonumber\\
&\left. +\frac{1}{(-\text{sgn}(t_2-t_3)\epsilon +i(t_2-t_3))^2(-\epsilon-i(t_3-a))(-\epsilon+i(t-t_2+a))} \right)  \nonumber\\
&=-\frac{\pi}{a^2}\int_{-\infty}^\infty \!\!\!\! dt dt_2\cos(\omega t)\cos(\omega_0t-\omega_1 t_2 )\left( \frac{1}{(\epsilon+it)^2}+ \frac{1}{(\epsilon-it)^2} \right)\left( \frac{1}{(\epsilon-it_2)^2(\epsilon-i(t_2+t))}+\frac{1}{(\epsilon+it_2)^2(\epsilon+i(t_2+t))} \right) \nonumber\\
&=\frac{2\pi^2}{a^2}\int_{-\infty}^\infty\!\!\!\! dt \cos(\omega t)\left( \frac{1}{(\epsilon+it)^2}+ \frac{1}{(\epsilon-it)^2} \right)\frac{1}{t^2}\left(  \cos((\omega_0+\omega_1)t) -\cos(\omega_0 t)+t\omega_1\sin(\omega_0 t) \right)        \nonumber\\
&=\frac{\pi^3}{3a^2} \left( -\left( |\omega+\omega_0+\omega_1|^3+ |\omega-\omega_0-\omega_1|^3\right) +\left( |\omega+\omega_0|^3+ |\omega-\omega_0|^3\right) +3\omega_1\left( \frac{|\omega+\omega_0|^2}{\text{sgn}(\omega+\omega_0)} - \frac{|\omega-\omega_0|^2}{\text{sgn}(\omega-\omega_0)} \right)  \right) \nonumber\\
&\xrightarrow{\omega=0}\frac{2\pi^3}{3a^2}\left( -|\omega_0+\omega_1|^3+|\omega_0|^3+3\omega_1\text{sgn}(\omega_0)|\omega_0|^2 \right) \nonumber\\
&\xrightarrow{\omega=0}\begin{cases}
-\frac{2\pi^3}{3a^2}|\omega_1|^3  & |\omega_0| \ll |\omega_1| \\
-\frac{2\pi^3}{a^2}|\omega_0|\omega_1^2 & |\omega_0| \gg |\omega_1|
\end{cases}
\end{align}
\end{widetext}
We assumed $a<0$ and since the result is symmetry under $a \rightarrow -a$, it holds for $a>0$ as well. 

From these results, we see that in any of the limits $|\omega_0|\gg|\omega_1| \gg 1/a$, $|\omega_1|\gg|\omega_0| \gg 1/a$, $|\omega_0|\ll |\omega_1| \ll 1/a$, or $|\omega_1|\ll |\omega_0| \ll 1/a$, the excess noise and excess current are proportional via $\Delta S_{tun}(\omega=0) = e\Delta I_{tun}$.

\section{Bosonic excess current and noise calculation} 
\label{sec:boson-calc}

Here we will show the details of how to find $\Delta I_{tun}$ to leading order as written in Eq~(\ref{eq:define-corr-function-boson}) using the correlators and Klein factors described in Appendix~\ref{sec:Keldysh} and the tunneling terms in Eq~(\ref{eq:L-tun-boson}) and (\ref{eq:L-inj-boson}). We first find the leading order current and noise, which follow similarly to the fermion case of the previous section. For the remainder of this section, we will omit the sum over quasiparticles and consider the contribution to the tunneling noise and current from a single species described by $q=(q_n,q_c)$ tunneling from the top edge at $x=0$ to a single species $q'=(q'_n,q'_c=q_c)$ at the bottom edge, with amplitude $\lambda$ and a single charge-$e$ species tunneling to the external lead described by $r=(r_n,r_c)$ with tunneling amplitude $\Lambda$. 
\begin{align}
&\langle I_{tun} \rangle_0 = \langle I_{tun} i\int dt \mathcal{L}_{tun} \rangle \nonumber\\
&= q_ce^*|\lambda|^2 \int_K dt \langle e^{iq_n\phi_{n,T}(0^+)}e^{-iq_n\phi_{n,T}(t^\mu)}\rangle\nonumber\\
&\times   \langle e^{iq_c\phi_{c,T}(0^+)}e^{-iq_c\phi_{c,T}(t^\mu)}\rangle \langle e^{-iq'_n\phi_{n,B}(0^+)}e^{iq'_n\phi_{n,B}(t^\mu)}\rangle \nonumber\\
&\times \langle e^{-iq'_c\phi_{c,B}(0^+)}e^{iq'_c\phi_{c,B}(t^\mu)}\rangle e^{-iq_c \omega_0t} + h.c. \nonumber\\
&= q_ce^*|\lambda|^2 \int_K dt \frac{-2i\sin(q_c \omega_0t)}{(\epsilon+iK^{+\mu}(-t)(-t))^{q^2+q'^2}} \nonumber\\
&= q_ce^*|\lambda|^2 \int_{-\infty}^0 dt \left( -2i\sin(q_c \omega_0t) \right) \left( \frac{1}{(\epsilon-it)^{q^2+q'^2}} -  \frac{1}{(\epsilon+it)^{q^2+q'^2}}  \right) \nonumber\\
&= \frac{2\pi q_ce^* |\lambda|^2 \text{sgn}(\omega_0)  |q_c\omega_0|^{q^2+q'^2-1}}{\Gamma(q^2+q'^2)}
\end{align}
The generalization from Eq~(\ref{eq:I-tun-0-Keldysh}) for fermions is clear. A similar calculation yields:
\begin{equation} \langle I_{inj}\rangle_0 = \frac{2\pi e|\Lambda|^2 \text{sgn}(\omega_1)|\omega_1|^{r^2}}{\Gamma(1+r^2)} \end{equation}

The leading order contribution to the shot noise at the QPC at $x=0$ is found in the same way, yielding
\begin{align}
S_{tun}(\omega) &= \frac{\pi |\lambda|^2 (q_ce^*)^2}{\Gamma(q^2+q'^2)}\left( |\omega+q_c\omega_0|^{q^2+q'^2-1} + |\omega-q_c\omega_0|^{q^2+q'^2-1} \right) \nonumber\\
&\xrightarrow{\omega=0} \frac{2\pi |\lambda|^2 (q_ce^*)^2}{\Gamma(q^2+q'^2)}|q_c\omega_0|^{q^2+q'^2-1}
\end{align}
This yields the expected proportionality $S_{tun}(\omega=0)= q_ce^*I_{tun}$. However, if there are multiple species of tunneling quasiparticles, then the total current and shot noise are a sum over the contributions from all quasiparticles and their proportionality is lost.

\subsection{Finite temperature}
\label{sec:finite-temp}
At finite temperature the correlation functions can be deduced from the zero-temperature correlators by conformal transformation. The result is\cite{Wen91}
\begin{equation}G^{\mu\nu}(t,x) \rightarrow \frac{\sin(\pi T G^{\mu\nu}(t,x))}{\pi T} \end{equation}
where $T$ indicates the temperature. Consequently, the results of the previous section are modified as follows:
\begin{align}
\langle I_{tun}\rangle_0 &= e^*|\lambda|^2 \int_{-\infty}^\infty dt \frac{-2i\sin(q_c \omega_0t)(\pi T)^{q^2 + q'^2}}{(\sin(\pi T(\epsilon-it))^{q^2 + q'^2}} \nonumber\\
&= e^*|\lambda|^2 (2\pi T)^{q^2 + q'^2-1}2i\sin \left(\frac{\pi}{2}\left( q^2 + q'^2 \right)\right)\nonumber\\
&\quad\quad \times B\left(1-q^2 + q'^2,-\frac{iq_c\omega_0}{2\pi T}+\frac{q^2+q'^2}{2} \right)+ h.c.
\end{align}
where $B$ is the beta-function $B(x,y)=\Gamma(x)\Gamma(y)/\Gamma(x+y)$. Similarly,
\begin{align} S_{tun}&(\omega=0) = (e^*)^2|\lambda|^2 (2\pi T)^{q^2 + q'^2-1}2\cos \left(\frac{\pi}{2} \left( q^2+q'^2 \right)\right)\nonumber\\
& \times \left( B\left(1-q^2 + q'^2, -\frac{iq_c\omega_0}{2\pi T}+\frac{q^2+q'^2}{2} \right) + h.c. \right)
\end{align}
There will be an additional contribution from interactions between the QPC and the noise in the source current.

\subsection{Excess current}
\label{sec:boson-current}
We now compute the correction to $\Delta I_{tun}$ from the injected current. This is similar to the calculation in Appendix~\ref{sec:fermion-excess-current}, but more difficult because the bosonized edge allows for fractional exponents. Expanding on the definition of $\Delta I_{tun}$ in Eq~(\ref{eq:define-corr-function-boson}) we find
\begin{align} &\frac{\Delta I_{tun}}{q_ce^*|\lambda|^2|\Lambda|^2 }=\nonumber\\
&  2\int_K \!\!\!\! dt_1^\mu dt_2^\nu dt_3^\sigma \frac{2i\sin(q_c \omega_0t_1 + \omega_1(t_3-t_2))}{(G^{+\mu}(-t_1,0))^{q^2 + q'^2}\left( G^{\nu\sigma}(t_2-t_3,0)\right)^{1+r^2}} H_{1R}H_{2L} 
\label{eq:delta-I-tun-Keldysh-boson}
\end{align}
where
\begin{align}
 &H_{j,R/L} = \frac{((-1)^j i)^{q_jr_j\left( K^{+\sigma}(-t_3)+K^{\mu\nu}(t_1-t_2)\right)} }{((-1)^j i)^{q_jr_j \left( K^{+\nu}(-t_2) + K^{\mu\sigma}(t_1-t_3) \right)}} \nonumber\\
&\times  \frac{\left( G^{+\nu}_{R/L}(-t_2,-a) \right)^{q_jr_j} \left( G^{\mu\sigma}_{R/L}(t_1-t_3,-a)\right)^{q_jr_j}}{ \left( G_{R/L}^{+\sigma}(-t_3,-a) \right)^{q_jr_j} \left( G_{R/L}^{\mu\nu}(t_1-t_2,-a) \right)^{q_jr_j}} \nonumber\\
&= \frac{ (i\epsilon K^{+\nu}(-t_2)-(-t_2 \pm a))^{q_jr_j}}{ (i\epsilon K^{+\sigma}(-t_3)-(-t_3 \pm a))^{q_jr_j}} \nonumber\\
&\times \frac{ (i\epsilon K^{\mu\sigma}(t_1-t_3)-(t_1-t_3 \pm a))^{q_jr_j}}{(i\epsilon K^{\mu\nu}(t_1-t_2)-(t_1-t_2 \pm a))^{q_jr_j}}
\end{align}
the index $j=1,2$ corresponds to $n,c$. The $R/L$ index has been suppressed on the correlation functions that have no spatial argument. The powers of $i$ in $H_{j,R/L}$ keep track of the Klein factors, as discussed in Appendix~\ref{sec:Keldysh}. When we sum over both sides of the Keldysh contour for the times $t_i$, we see that only times $t_i <0$ survive; this is another example of the Keldysh method enforcing causality. Hence, the integral (\ref{eq:delta-I-tun-Keldysh-boson}) is dominated by $t_1 \approx 0$, $t_{2,3}\approx -|a|$. As shown in Fig~\ref{fig:ExptSetup}, we are interested in $a<0$, although by the symmetry of Eq~(\ref{eq:delta-I-tun-Keldysh-boson}), the computation for $a>0$ will be the same as that for $a<0$ if we swap $q_n,r_n \leftrightarrow q_c,r_c$. Consequently, we are free to take $a<0$. In this case, the branch cuts in $H_{2L}$ do not approach zero, so we can take $\epsilon=0$ in $H_{2L}$, as well as $t_2=t_3$, and find $H_{2L}=1$. We cannot make these approximations in $H_{1R}$ because $\epsilon$ will matter when the branch cuts get close to zero. Writing out the sum over all parts of the Keldysh contour with this simplification yields
\begin{widetext}
\begin{align}
 \frac{\Delta I_{tun}}{q_ce^*|\lambda|^2|\Lambda|^2} &= 4i \int_{-\infty}^0 dt_1dt_2dt_3 \frac{\sin(q_c \omega_0t_1 + \omega_1(t_3-t_2))}{(\epsilon-it_1)^{q^2 + q'^2}} \nonumber\\
 &\times \left[ \frac{1}{(\epsilon+i|t_2-t_3|)^{1+r^2}}\left( \frac{(i\epsilon-(-t_2+a))(i\epsilon\text{sgn}(t_1-t_3)-(t_1-t_3+a))}{(i\epsilon-(-t_3+a))(i\epsilon\text{sgn}(t_1-t_2)-(t_1-t_2+a))} \right)^{q_nr_n}\right. \nonumber\\
 & - \frac{1}{(\epsilon-i(t_2-t_3))^{1+r^2}}\left( \frac{(i\epsilon-(-t_2+a))(-i\epsilon-(t_1-t_3+a))}{(-i\epsilon-(-t_3+a))(i\epsilon\text{sgn}(t_1-t_2)-(t_1-t_2+a))} \right)^{q_nr_n} \nonumber\\
 & - \frac{1}{(\epsilon+i(t_2-t_3))^{1+r^2}}\left( \frac{(-i\epsilon-(-t_2+a))(i\epsilon\text{sgn}(t_1-t_3)-(t_1-t_3+a))}{(i\epsilon-(-t_3+a))(-i\epsilon-(t_1-t_2+a))} \right)^{q_nr_n} \nonumber\\
 & \left.  +\frac{1}{(\epsilon-i|t_2-t_3|)^{1+r^2}}\left( \frac{(-i\epsilon-(-t_2+a))(-i\epsilon-(t_1-t_3+a))}{(-i\epsilon-(-t_3+a))(-i\epsilon-(t_1-t_2+a))} \right)^{q_nr_n} \right] + h.c. \nonumber\\
%
%
&= 4i \int_{-\infty}^0 dt_1dt_2dt_3 \sin(q_c \omega_0t_1 + \omega_1(t_3-t_2))\left(\frac{1}{(\epsilon-it_1)^{q^2 + q'^2}} - \frac{1}{(\epsilon+it_1)^{q^2 + q'^2}} \right)\nonumber\\
&\times \left[ \Theta(t_2-t_3)\left( \frac{1}{(\epsilon+i(t_2-t_3))^{1+r^2}}\frac{(i\epsilon-(t_1-t_3+a))^{q_nr_n}}{(i\epsilon-(-t_3+a))^{q_nr_n}} - \frac{1}{(\epsilon-i(t_2-t_3))^{1+r^2}}\frac{(-i\epsilon-(t_1-t_3+a))^{q_nr_n}}{(-i\epsilon-(-t_3+a))^{q_nr_n}} \right)  \right. \nonumber\\
&\quad\quad\times   \left( \frac{(i\epsilon-(-t_2+a))^{q_nr_n}}{(i\epsilon-(t_1-t_2+a))^{q_nr_n}} - \frac{(-i\epsilon-(-t_2+a))^{q_nr_n}}{(-i\epsilon-(t_1-t_2+a))^{q_nr_n}}\right) \nonumber\\
&+\Theta(t_3-t_2) \left( \frac{1}{(\epsilon-i(t_2-t_3))^{1+r^2}}\frac{(i\epsilon-(-t_2+a))^{q_nr_n}}{(i\epsilon-(t_1-t_2+a))^{q_nr_n}} -  \frac{1}{(\epsilon+i(t_2-t_3))^{1+r^2}}\frac{(-i\epsilon-(-t_2+a))^{q_nr_n}}{(-i\epsilon-(t_1-t_2+a))^{q_nr_n}} \right) \nonumber\\
&\quad\quad \times \left.\left( \frac{(i\epsilon-(t_1-t_3+a))^{q_nr_n}}{(i\epsilon-(-t_3+a))^{q_nr_n}} - \frac{(-i\epsilon-(t_1-t_3+a))^{q_nr_n}}{(-i\epsilon-(-t_3+a))^{q_nr_n}}\right) \right] \nonumber\\
%
%
&= 4i \int_{-\infty}^\infty dt_1 \int_{-\infty}^0 dt_2dt_3 \sin(q_c \omega_0t_1 + \omega_1(t_3-t_2))\left(\frac{1}{(\epsilon-it_1)^{q^2 + q'^2}} - \frac{1}{(\epsilon+it_1)^{q^2 + q'^2}} \right)  \Theta(t_2-t_3)  \nonumber\\
&\times \left( \frac{1}{(\epsilon+i(t_2-t_3))^{1+r^2}}\frac{(i\epsilon-(t_1-t_3+a))^{q_nr_n}}{(i\epsilon-(-t_3+a))^{q_nr_n}}-  \frac{1}{(\epsilon-i(t_2-t_3))^{1+r^2}}\frac{(-i\epsilon-(t_1-t_3+a))^{q_nr_n}}{(-i\epsilon-(-t_3+a))^{q_nr_n}} \right)   \nonumber\\
&\quad\quad\times   \left( \frac{(i\epsilon-(-t_2+a))^{q_nr_n}}{(i\epsilon-(t_1-t_2+a))^{q_nr_n}} - \frac{(-i\epsilon-(-t_2+a))^{q_nr_n}}{(-i\epsilon-(t_1-t_2+a))^{q_nr_n}}\right) \nonumber\\
%
%
&= 4i \int_{-\infty}^\infty dt_1 dt_2dt_3 \sin(q_c \omega_0t_1 + \omega_1(t_3-t_2))\left(\frac{1}{(\epsilon-it_1)^{q^2 + q'^2}} - \frac{1}{(\epsilon+it_1)^{q^2 + q'^2}} \right)  \Theta(t_2-t_3)  \nonumber\\
&\times \left( \frac{1}{(\epsilon+i(t_2-t_3))^{1+r^2}}\frac{(i\epsilon-(t_1-t_3))^{q_nr_n}}{(i\epsilon-(-t_3))^{q_nr_n}}-  \frac{1}{(\epsilon-i(t_2-t_3))^{1+r^2}}\frac{(-i\epsilon-(t_1-t_3))^{q_nr_n}}{(-i\epsilon-(-t_3))^{q_nr_n}} \right)   \nonumber\\
&\quad\quad\times   \left( \frac{(i\epsilon-(-t_2))^{q_nr_n}}{(i\epsilon-(t_1-t_2))^{q_nr_n}} - \frac{(-i\epsilon-(-t_2))^{q_nr_n}}{(-i\epsilon-(t_1-t_2))^{q_nr_n}}\right)
\label{eq:I-tun-boson-manipulations}\end{align}
\end{widetext}
Notice that we have replaced $\text{sgn}(t_1-t_{2,3})$ with $1$; this should not change the integral since it has little or no contribution when $t_1 < t_{2,3}$. To get the third equality, in the term involving $\Theta(t_3-t_2)$ we shift $t_{2,3} \rightarrow t_{2,3}+t_1$, then take $t_1 \rightarrow -t_1$, $t_2 \leftrightarrow t_3$. This procedure produces an extra region of integration, $t_1>0$, $0<t_{2,3}<t_1$, but the integral over this region is negligible. To obtain the fourth equality we shift $t_{2,3} \rightarrow t_{2,3} + a$ and add another negligible region of integration so that all $t_i$ have the same limits of integration. We have assumed $|a| \gg 1/|\omega_0|,1/|\omega_1|$.

The remaining integral is difficult to do analytically, but
by rescaling the variables it is evident that $\Delta I_{tun}$ will be a sum of terms $|\omega_0|^\alpha|\omega_1|^\beta$ where $\alpha+\beta = q^2 + q'^2+r^2-2$. We expect that exponents $\alpha,\beta$ that appear will be independent of the product $q_nr_n$, although the coefficients might be dependent. To make progress on the integral above, we here consider $q_nr_n=1$, knowing that it will probably give us the right terms with the wrong coefficients; we have confirmed this intuition by repeating the computation with $q_nr_n=2$ (not shown). With this simplification, 
\begin{widetext}
\begin{align}
\frac{\Delta I_{tun}}{q_ce^*|\lambda|^2|\Lambda|^2}&= 4\pi \int_{-\infty}^\infty dt_1 dt_3 \sin(q_c \omega_0t_1 +\omega_1(t_3-t_1))\left( \frac{1}{(\epsilon-it_1)^{q^2 + q'^2}}-\frac{1}{(\epsilon+it_1)^{q^2 + q'^2}}\right)(t_1)\nonumber\\
&\times \left( \frac{i}{(\epsilon+i(t_1-t_3))^{r^2}(t_3+i\epsilon)} + \frac{i}{(\epsilon-i(t_1-t_3))^{r^2}(t_3-i\epsilon)} \right)
\end{align}
\end{widetext}
We have utilized the delta-function identity (\ref{eq:delta-function}). If we assume $r^2$ is an integer, we can do the integral over $t_3$ as a contour integral, as shown in Eq~(\ref{eq:useful-integral})), and then assume analytic continuation to all $n$. The result is
\begin{widetext}
\begin{align}
& \frac{\Delta I_{tun}}{q_ce^*|\lambda|^2|\Lambda|^2} = -\text{sgn}(\omega_1)2\pi \int_{-\infty}^\infty dt_1 \left( \frac{t_1}{(\epsilon-it_1)^{q^2 + q'^2}}-\frac{t_1}{(\epsilon+it_1)^{q^2 + q'^2}}\right) \int_{-\infty}^\infty dt_3 \left( \frac{e^{-i\text{sgn}(\omega_1)(q_c \omega_0 t_1+\omega_1(t_3-t_1))}}{(\epsilon+i(t_1-t_3))^{r^2}(t_3+i\epsilon)}  \right)+h.c. \nonumber\\
&= \text{sgn}(\omega_1)\frac{8\pi^2(-i)^{r^2-1}}{\Gamma(r^2)}\int_{0}^\infty dt_1\left( \frac{1}{(\epsilon-it_1)^{q^2 + q'^2}}-\frac{1}{(\epsilon+it_1)^{q^2 + q'^2}}\right)  \frac{\left(\Gamma(r^2)-\Gamma(r^2,it_1|\omega_1|)\right)e^{it_1\text{sgn}(\omega_1)(\omega_1-q_c\omega_0)}}{t_1^{r^2-1}} + h.c. \nonumber\\
&= \text{sgn}(\omega_1)\frac{16\pi^2(-i)^{r^2-2}\sin(\frac{1}{2}\pi(q^2+q'^2))}{\Gamma(r^2)}\int_{0}^\infty dt_1 \frac{\left(\Gamma(r^2)-\Gamma(r^2,it_1|\omega_1|)\right)e^{it_1\text{sgn}(\omega_1)(\omega_1-q_c\omega_0)}}{t_1^{r^2+q^2 + q'^2-1}} + h.c.  \nonumber\\
&= \text{sgn}(\omega_1)16\pi^2(i)^{q^2 + q'^2} \sin(\frac{1}{2}\pi(q^2+q'^2))|q_c\omega_0|^\Delta \left(\text{sgn}(\omega_1\omega_0)\right)^{\Delta} \left( \left(1-\frac{\omega_1}{q_c\omega_0}\right)^{\Delta}\Gamma(-\Delta)-\sum_{k=0}^{r^2-1}\frac{1}{k!}\left(\frac{\omega_1}{q_c\omega_0}\right)^k\Gamma(-\Delta+k) \right) + h.c. \nonumber\\
&=\begin{cases}
 \text{sgn}(\omega_1)16\pi^2 \sin( \pi (q^2+q'^2))|q_c\omega_0|^{q^2 + q'^2-2}|\omega_1|^{r^2} \displaystyle\sum_{k=0}^\infty \frac{\Gamma(k-q^2 - q'^2+2)}{\Gamma(k+r^2+1)}\left( \frac{\omega_1}{q_c\omega_0}\right)^{k} \text{, if }|\omega_0| \gg |\omega_1| \\
\text{sgn}(\omega_1)16\pi^2 \sin( \pi (q^2+q'^2))|\omega_1|^\Delta \!\! \left(\displaystyle\sum_{k=0}^\infty \frac{\Gamma(k-\Delta)}{\Gamma(k+1)}(-1)^{r^2} \!\! \left(\frac{q_c\omega_0}{\omega_1}\right)^k \!\! -\!\! \displaystyle\sum_{k=0}^{r^2-1} \frac{\Gamma(1-q^2 - q'^2-k)}{\Gamma(r^2-k)}\left| \frac{q_c\omega_0}{\omega_1}\right|^{q^2 + q'^2-1+k}\!\!\!\! \text{sgn}(\omega_0\omega_1)^{k+1}\right), \\
\quad\quad \text{if }|\omega_1| \gg |\omega_0| 
 \end{cases}
\label{eq:Delta-I-tun-integral}\end{align}
\end{widetext}
where we have defined $\Delta = r^2+q^2 + q'^2-2$ and assumed $n \in \mathbb{Z}$.
Notice that this reproduces the results of Eq~(\ref{eq:tunn-current-fermions-appendix-0}) in either limiting case. 

This calculation gives the contribution to the change in tunneling current from a single species $q = (q_n,q_c)$ tunneling from the top edge of the Hall bar to a species $q'=(q'_n,q'_c)$ on the bottom edge and a single species $r=(r_n,r_c)$ tunneling into the external lead. Physically, quasiparticles with $q_n \rightarrow -q_n$ and $r_n\rightarrow -r_n$ will also be present and could tunnel from the top edge to a species $q'$ on the bottom edge. From the symmetry of the model, taking $q_nr_n \rightarrow -q_nr_n$ is equivalent to taking $\omega_1 \rightarrow -\omega_1$ in the computation of $\Delta I_{tun}$. Hence, if the tunneling amplitudes for the two types of quasiparticles with opposite contributions from the neutral mode are equal, then the leading contributions from Eq~({\ref{eq:Delta-I-tun-integral}) will cancel and the subleading terms will dominate. In this case, $\Delta I_{tun}$ will be even in $\omega_1$ and odd in $\omega_0$. 

When $a \rightarrow -a$, i.e. the current injection is upstream of the QPC, $q_nr_n \rightarrow q_cr_c $ in Eq~(\ref{eq:I-tun-boson-manipulations}). At first, this transformation seems inconsequential -- after all, we argued that this exponent only changes the final answer by an overall pre-factor -- but it becomes important when regarding the symmetry considerations of the previous paragraph. Namely, when $a>0$, Eq~(\ref{eq:I-tun-boson-manipulations}) is invariant under $q_n \rightarrow -q_n$. Consequently, when the contributions from quasiparticles with $q=(q_n,q_c)$ and $(-q_n,q_c)$ are added, $\Delta I_{tun}$ doubles, in contrast to the case when $a<0$ and terms odd in $\omega_1$ disappear. If $q^2>1/2$, the leading term in $\Delta I_{tun}$ is odd in $\omega_1$; hence, when $a>0$, $\Delta I_{tun}$ is larger by a power of $\rm{Max}\left( \frac{\omega_0}{\omega_1},\frac{\omega_1}{\omega_0}\right)$ than when $a<0$. This agrees with the intuition that there should be a larger change in tunneling current when the injection is upstream of the QPC than when it is downstream.

\subsection{Excess noise}

Using Eq~(\ref{eq:L-tun-boson}) and (\ref{eq:L-inj-boson}), the excess noise is given by
\begin{align}
\Delta &S_{tun}(t) = \frac{1}{2} \left( \langle \lbrace I_{tun}(t), I_{tun}(0) \rbrace \rangle |_{\Lambda} - \langle \lbrace I_{tun}(t), I_{tun}(0) \rbrace \rangle |_{\Lambda=0} \right) \nonumber\\
&=\frac{1}{2} \langle \lbrace I_{tun}(t), I_{tun}(0) \rbrace \left( i\int dt_2 \mathcal{L}_{inj} \right) \left( i\int dt_3\mathcal{L}_{inj} \right) \rangle_0 
\end{align}
Here we seek $\Delta S_{tun}(\omega) = \int dt e^{i\omega t}\Delta S_{tun}(t)$. The correlation functions and Klein factors of Appendix~\ref{sec:Keldysh} yield the contribution from a single pair of quasiparticle species described by $m=(q_n,q_c), n=(r_n,r_c)$,
\begin{align}
&\frac{\Delta S_{tun}}{(q_ce^*)^2|\lambda|^2|\Lambda|^2} = - \int _{-\infty}^\infty dt 2\cos(\omega t) \int_K dt_2^\nu dt_3^\sigma \nonumber\\
&\quad\quad \frac{2\cos(q_c \omega_0t + \omega_1(t_3-t_2))}{G^{+-}(-t_1,0))^{q^2 + q'^2}(G^{\nu\sigma}(t_2-t_3))^{1+r^2}}H_{1R}H_{2L}
\end{align}
where $H_{iR/L}$ are defined with $\mu = -$. The similarity to Eq~(\ref{eq:delta-I-tun-Keldysh-boson}) is clear. Following the manipulations of Eq~(\ref{eq:I-tun-boson-manipulations}), 
\begin{widetext} \begin{align}
\frac{\Delta S_{tun}(\omega)}{(q_ce^*)^2|\lambda|^2|\Lambda|^2}
&= -4 \int_{-\infty}^\infty dt dt_2dt_3 \cos(\omega t) \cos(q_c \omega_0t+ \omega_1(t_3-t_2))\left(\frac{1}{(\epsilon-it)^{q^2 + q'^2}} + \frac{1}{(\epsilon+it)^{q^2 + q'^2}} \right)  \Theta(t_2-t_3)  \nonumber\\
&\times \left( \frac{1}{(\epsilon+i(t_2-t_3))^{1+r^2}}\frac{(i\epsilon-(t-t_3))^{q_nr_n}}{(i\epsilon-(-t_3))^{q_nr_n}} - \frac{1}{(\epsilon-i(t_2-t_3))^{1+r^2}}\frac{(-i\epsilon-(t-t_3))^{q_nr_n}}{(-i\epsilon-(-t_3))^{q_nr_n}} \right)   \nonumber\\
&\quad\quad\times   \left( \frac{(i\epsilon-(-t_2))^{q_nr_n}}{(i\epsilon-(t-t_2))^{q_nr_n}} - \frac{(-i\epsilon-(-t_2))^{q_nr_n}}{(-i\epsilon-(t-t_2))^{q_nr_n}}\right)
\label{eq:delta-S-tun-integral-0}
\end{align}
\end{widetext}
We now make the simplifying assumption that $q_nr_n=1$; as in the current case, we have separately checked that when $q_nr_n=2$, the only change is a pre-factor (which is the same pre-factor as in the current case). Note that in this case, though, the transformation $q_nr_n\rightarrow -q_nr_n$ is equivalent to $ \omega_0\rightarrow -\omega_0$. Under this assumption, 
\begin{widetext}
\begin{align}
\frac{\Delta S_{tun}(\omega)}{(q_ce^*)^2|\lambda|^2|\Lambda|^2}
&= 4\pi i \int_{-\infty}^\infty dt dt_3 \cos(\omega t)\cos(q_c \omega_0t + \omega_1(t_3-t))\left( \frac{1}{(\epsilon-it)^{q^2 + q'^2}} + \frac{1}{(\epsilon+it)^{q^2 + q'^2}} \right)(t) \nonumber\\
&\times \left( \frac{i}{(\epsilon+i(t-t_3))^{r^2}(t_3+i\epsilon)} + \frac{i}{(\epsilon-i(t-t_3))^{r^2}(t_3-i\epsilon)} \right) \nonumber\\
&= -2\pi  \int_{-\infty}^\infty dt dt_3 \cos(\omega t)\left( \frac{t}{(\epsilon-it)^{q^2 + q'^2}} + \frac{t}{(\epsilon+it)^{q^2 + q'^2}} \right) \left( \frac{e^{-i\text{sgn}(\omega_1)(q_c\omega_0t +\omega_1(t_3-t)}}{(\epsilon+i(t-t_3))^{r^2}(t_3+i\epsilon)} \right) + h.c. \nonumber\\
&= \frac{8\pi^2 (-i)^{r^2-1}}{\Gamma(r^2)}  \int_{0}^\infty dt  \cos(\omega t)\left( \frac{1}{(\epsilon-it)^{q^2 + q'^2}} + \frac{1}{(\epsilon+it)^{q^2 + q'^2}} \right)\frac{\left( (\Gamma(r^2)-\Gamma(r^2,it|\omega_1|) \right)e^{it(|\omega_1|-q_c \omega_0\text{sgn}(\omega_1))}}{t^{r^2-1}} + h.c. \nonumber\\
&= \frac{16\pi^2 (-i)^{r^2-1}\cos(\frac{\pi}{2} \left(q^2+ q'^2\right))}{\Gamma(r^2)}  \int_{0}^\infty dt  \cos(\omega t)\frac{\left( (\Gamma(r^2)-\Gamma(r^2,it|\omega_1|) \right)e^{it(|\omega_1|- q_c \omega_0\text{sgn}(\omega_1))}}{t^{q^2 + q'^2+r^2-1}} + h.c. \nonumber\\
&\xrightarrow{\omega \rightarrow 0}\begin{cases}
\text{sgn}(\omega_1\omega_0)16\pi^2 \sin( \pi (q^2+q'^2))|q_c\omega_0|^{q^2 + q'^2-2}|\omega_1|^{r^2} \displaystyle\sum_{k=0}^\infty \frac{\Gamma(k-q^2 - q'^2+2)}{\Gamma(k+r^2+1)}\left( \frac{\omega_1}{q_c\omega_0}\right)^{k} \text{, if }|\omega_0| \gg |\omega_1| \\
-16\pi^2 \sin(\pi (q^2+q'^2) )|\omega_1|^\Delta \left( \displaystyle\sum_{k=0}^\infty \frac{\Gamma(k-\Delta)}{\Gamma(k+1)}(-1)^{r^2} \!\! \left(\frac{q_c\omega_0}{\omega_1}\right)^{\!\! k}  \right. \\
\quad\quad\quad\quad\quad\quad\quad\quad\quad \left.  - \displaystyle\sum_{k=0}^{r^2-1} \frac{\Gamma(1-q^2 - q'^2-k)}{\Gamma(r^2-k)}\left| \frac{q_c\omega_0}{\omega_1}\right|^{q^2 + q'^2-1+k} \!\!\!\! \text{sgn}(\omega_0\omega_1)^{k+1}\right) \text{, } \text{if }|\omega_1| \gg |\omega_0| \\
\end{cases}
\label{eq:delta-S-tun-integral}\end{align}
\end{widetext}
We have utilized the delta-function identity Eq~(\ref{eq:delta-function}) and the integral Eq~(\ref{eq:useful-integral}), taken $n \in \mathbb{Z}$ and defined $\Delta = r^2+q^2 + q'^2-2$. The excess noise from a single species $q=(q_n,q_c)$ on the top edge tunneling across the QPC to a species $q'=(q'_n,q'_c)$ on the bottom edge and a species $r=(r_n,r_c)$ tunneling from the external lead is proportional to the contribution to the excess current from the same species. When quasiparticles with $q_n \rightarrow -q_n$ and $r_n \rightarrow -r_n$ are also present and tunnel with equal amplitudes, their contribution to the noise will be given by Eq~(\ref{eq:delta-S-tun-integral}) with $\omega_0\rightarrow -\omega_0$. 

Now consider the case of $a>0$. Similar to the discussion at the end of the previous section, when $a>0$, $q_nr_n \rightarrow q_cr_c$ in Eq~\ref{eq:delta-S-tun-integral-0}. Consequently, when $a>0$ and the contributions to the excess noise from quasiparticles with $(\pm q_n, q_c)$ are added together, the excess noise doubles. This is in contrast to the $a<0$ case when the terms odd in $\omega_0$ drop out. Thus, when $|\omega_0| \gg |\omega_1|$ or when $q^2 < 1/2$ and $|\omega_1| \gg |\omega_0|$, the excess noise will increase by a factor of $\text{Max}\left( \frac{\omega_0}{\omega_1},\frac{\omega_1}{\omega_0} \right)$ when $a>0$.

\subsection{A useful integral}
\begin{align}
& \int_{-\infty}^\infty dt_3 \frac{e^{-i(\omega_0\text{sgn}(\omega_1)-|\omega_1|)t_1-i|\omega_1|t_3}}{(t_3-t_1+i\epsilon)^{r^2}(t_3+i\epsilon)} \nonumber\\
&= -2\pi i \left( \frac{e^{-i(\omega_0\text{sgn}(\omega_1)-|\omega_1|)t_1}}{(-t_1)^{r^2}} \right.\nonumber\\
&\quad\quad \left. +  \sum_{k=0}^{r^2-1} \frac{(-1)^{r^2-1-k}(-i|\omega_1|)^k e^{-i\omega_0\text{sgn}(\omega_1)t_1}}{k! \, t_1^{r^2-k}}\right) \nonumber\\
&= -2\pi i \left( \sum_{k=0}^\infty \frac{(-1)^{r^2}\left(i|\omega_1|\right)^k}{k! \, t_1^{r^2-k}} \right. \nonumber\\
&\quad\quad  \left. + \sum_{k=0}^{r^2-1} \frac{(-1)^{r^2-1}(i|\omega_1|)^k }{k! \, t_1^{r^2-k}}\right)e^{-i\omega_0\text{sgn}(\omega_1)t_1} \nonumber\\
&=  -2\pi i \sum_{k=r^2}^\infty  \frac{(-1)^{r^2}\left(i|\omega_1|\right)^k}{k! \, t_1^{r^2-k}} e^{-i\omega_0\text{sgn}(\omega_1)t_1} \nonumber\\
&= -2\pi (-i)^{r^2} \sum_{k=0}^\infty \frac{(i)^{k+1}|\omega_1|^{k+r^2} }{(k+r^2)! }t_1^{k} e^{-i\omega_0\text{sgn}(\omega_1)t_1} \nonumber\\
&= -\frac{2\pi i(-1)^{r^2}}{\Gamma(r^2)t_1^{r^2}}\left(\Gamma(r^2)-\Gamma(r^2,it_1|\omega_1|) \right)e^{it_1(|\omega_1|-\omega_0\text{sgn}(\omega_1))}
\label{eq:useful-integral}
\end{align}
where $\Gamma(r^2,x)$ is the incomplete gamma function. The last equality gives the result for $n \not\in \mathbb{Z}$ by analytic continuation.

\subsection{Specific results for $r^2=2, q^2=2/3$}
\label{sec:compute23}
As discussed in the text, the $\nu=2/3$ edge is expected to be described by $r^2=2, q^2=2/3$. Using these values, we can do the integrals in Eqs~(\ref{eq:Delta-I-tun-integral}) and (\ref{eq:delta-S-tun-integral}) exactly (bear in mind that we expect these integrals to be correct up to a constant of proportionality, since we have assumed $q_nr_n=1$, which does not correctly describe the state, but should not be expected to change the scaling):
\begin{align}
\Delta & I_{tun}=\frac{e}{3}|\lambda|^2|\Lambda|^2 \frac{16\pi^3}{\Gamma(7/3)}\text{sgn}(\omega_1)\nonumber\\
&\times\left(-|q_c\omega_0-\omega_1|^{4/3}-\frac{4}{3}\omega_1\text{sgn}(\omega_0)|q_c\omega_0|^{1/3}+|q_c\omega_0|^{4/3}\right) \\
\Delta & S_{tun} = \left( \frac{e}{3} \right)^2|\lambda|^2|\Lambda|^2\frac{16\pi^3}{\Gamma(7/3)}\left( \text{sgn}(1-q_c\omega_0\omega_1)|q_c\omega_0-\omega_1|^{4/3}\right.\nonumber\\
&\left. -\frac{4}{3}|\omega_1||q_c\omega_0|^{1/3} + \text{sgn}(\omega_0\omega_1)|q_c\omega_0|^{4/3} \right)
\end{align}
where $q_c\omega_0 = eV_0/3$.
When we assume that quasiparticles with $q_n \rightarrow -q_n$ and $r_n \rightarrow -r_n$ tunnel with equal amplitudes, we obtain Eqs~(\ref{eq:excess-noise-symm}) and (\ref{eq:excess-current-symm}) in the main text.

\bibliography{UnexpectedCurrent}

\end{document}